\documentclass[aps,rmp,reprint,nofootinbib,
preprintnumbers,amsmath,amssymb,superscriptaddress,longbibliography]{revtex4-2}
\usepackage[utf8]{inputenc}
\usepackage{graphicx}
\usepackage{subfigure}
\usepackage{soul}
\usepackage{color}
\usepackage{changes}
\usepackage{changes}
\usepackage[english]{babel}
\usepackage[pdftex]{hyperref}
\hypersetup{ 
pdfstartview=FitV,
colorlinks=true, 
linkcolor=black, 
citecolor=blue, 
filecolor=black, 
urlcolor=magenta,
}

\newcommand{\bea}{\begin{eqnarray}}
\newcommand{\eea}{\end{eqnarray}}

\graphicspath{{Figures/}}

\begin{document}

\preprint{CPHT-PC014.022022}

\title{\Large \textit{Colloquium}: Hydrodynamics and holography of charge density wave phases}

%\date{\today}

\author{Matteo Baggioli}\email[]{b.matteo@sjtu.edu.cn}
\affiliation{Wilczek Quantum Center, School of Physics and Astronomy, Shanghai Jiao Tong University, Shanghai 200240, China \vspace{0.2cm}}
\affiliation{Shanghai Research Center for Quantum Sciences, Shanghai 201315, China\vspace{0.2cm}}

\author{Blaise Gout\'eraux \vspace{0.4cm}}\email[]{blaise.gouteraux@polytechnique.edu}

\affiliation{CPHT, CNRS, \'Ecole Polytechnique, IP Paris, F-91128 Palaiseau, France}

\begin{abstract}
In this {\em Colloquium}, we review recent progress in the effective description of strongly-correlated phases of matter with spontaneously broken translations, such as charge density waves or Wigner crystals. In real materials, disorder is inevitable and pins the Goldstones of broken translations. We describe how pinning can be incorporated in the effective field theory at low energies, without making any assumption on the presence of boost symmetry. We review the essential role played by gauge-gravity duality models in establishing these effective field theories with only approximate symmetries. We close with a discussion on the relevance of these models for the phenomenology of dc and ac transport in strongly-correlated strange and bad metals, such as high temperature superconductors.
\end{abstract}

\date{\today} 
\maketitle
\tableofcontents
%%%%%%%%%%%%%%%%%%%%%%
\section{Introduction}

Strongly-correlated states of matter present a serious theoretical challenge, as perturbation theory typically fails to describe them. High critical temperature ($T_c$) superconductors \cite{bednorz1986} constitute an archetypal example and have resisted theoretical efforts to account for their phenomenology since their discovery, \cite{keimer2015quantum,alexandradinata2020future}. The absence of long-lived quasiparticles, as reported by photo-emission experiments, and their unconventional transport properties are two signatures of their incompatibility with the Fermi liquid paradigm, \cite{landau1980course}. The Hubbard model \cite{arovas2021hubbard}, marginal Fermi liquid theory \cite{PhysRevLett.63.1996} and various field theories with a large (infinite) number of degrees of freedom \cite{sachdev2011quantum} aided by random interactions \cite{Chowdhury:2021qpy} all provide some degree of insight into this problem.

Progress in understanding the physics at play in these systems has been complicated by the variety of phases that appear to be competing (or working in concert) in different regions of the phase diagram characterized by temperature, doping, magnetic field, pressure, etc.

Hydrodynamics and effective field theory methods \cite{Kovtun:2012rj,Nicolis:2015sra,Glorioso:2018wxw} offer a complementary avenue by eschewing the microscopic details of strongly-correlated systems, as done for example in the cases of graphene \cite{Lucas:2017idv} and bad metals \cite{Hartnoll:2014lpa}.  The price to pay is that the analysis is limited to low energies, late times and long distances, and breaks down at very low temperatures (in particular in the vicinity of any quantum critical point where fluctuation effects cannot be neglected). The effective frameworks also take as input a number of parameters which are constrained by various consistency requirements but the values of which can only be computed within a microscopic completion.

Gauge-gravity duality (also referred to colloquially as holography) maps a strongly-coupled, large $N$ matrix model (where $N$ is the rank of the gauge group) to classical Einstein gravity coupled to a set of matter fields, \cite{Maldacena:1997re}. The application of this set of techniques to strongly-correlated condensed matter systems has been intensively pursued in the past fifteen years, \cite{zaanen2015holographic,Hartnoll:2016apf}. The original duality \cite{Maldacena:1997re} relates a specific gauge theory to a specific string theory, so that in principle microscopic degrees of freedom on both sides of the duality can be matched (in practice, this can be technically involved). A more common approach in applied gauge-gravity duality is the bottom up one, where the dual field theory is not known precisely, nor is it clear that the classical gravity dual can be promoted to a full quantum gravity. Assuming though that such a dual field theory exists, and within the range of validity of the low energy classical gravity theory, the equation of state and transport coefficients of its classical saddle point can be computed. 

In applying these tools, identifying the right set of symmetries is paramount, as this will dictate the starting point of the effective approach. This does so by determining the set of hydrodynamic conservation equations governing the low energy dynamics of the system in one case, or by acting as a guiding principle to write down the appropriate bulk action in the second. 

The aim of this \textit{Colloquium} is to describe recent progress in effective hydrodynamic and holographic theories of phases with spontaneously broken translations,  motivated by the ubiquity of such phases in the phase diagram of strongly-correlated electron materials, in particular cuprate or iron-based high $T_c$ superconductors, kagome materials, organic conductors, transition metals dichalcogenides, etc. While translational `spin-charge stripe' order was long anticipated on theoretical grounds to play an important role in underdoped cuprates and other doped Mott insulators \cite{PhysRevB.40.7391,MACHIDA1989192,PhysRevB.39.9749,Kivelson1998,PhysRevLett.108.267001,PhysRevB.86.115138,Beekman:2016szb}, and was experimentally confirmed subsequently in most families of underdoped cuprate materials \cite{Frano:2020} as well as in numerical studies of the Hubbard model \cite{Huang_2017,Zheng_2017}, recent experiments suggest that charge density fluctuations \cite{Kivelson:2003zz} and short-range charge density wave order are actually found across the phase diagram,  \cite{Peng:2018,Arpaia:2019,Miao:2021,Lin:2021,Lee:2020,Ma:2021,Arpaia:2021,Tam:2021,Kawasaki:2021,Lee:2021generic}.\footnote{A recent numerical study of the Hubbard model also reports fluctuating stripes across the phase diagram, \cite{Huang:2022}.} These observations prompted a number of investigations of the impact of such fluctuating charge order on transport and spectroscopic experiments, \cite{PhysRevB.95.224511,Delacretaz:2016ivq,Delacretaz:2017zxd,Amoretti:2018tzw,Seibold:2021,Delacretaz:2021qqu}.

In seeking to apply effective field theory methods to this problem, one is inevitably confronted with the impact of disorder and other sources of explicit translation symmetry breaking on the dynamics of the charge density wave, leading to the phenomenon of pinning, \cite{RevModPhys.60.1129}. When the explicit breaking is weak, the symmetry rules which usually tightly constrain effective field theories are relaxed and it becomes more arduous to develop a consistent double expansion, in powers of the strength of the explicit breaking and of the effective field theory cut-off. On the other hand, gauge-gravity duality allows to model such phases from first principles, and acts as a testing arena for effective field theories with approximate symmetries.

The first two sections \ref{sec1}-\ref{sec2} give a brief review of hydrodynamics and of holographic methods. In section \ref{sec11}, we then describe recent progress in incorporating background strain in the hydrodynamics of spontaneously broken translation phases, without assuming any particular boost symmetry, and expand on verifications of this theory by various holographic models. Next, in section \ref{sec:3}, we turn to the physics of pseudo-spontaneous translation symmetry breaking in hydrodynamics and in holography, and discuss phenomenological implications. We also comment on the role of topological defects and magnetic fields. 

\section{Hydrodynamics}
\label{sec1}

Hydrodynamics \cite{KADANOFF1963419,forster,chaikin2000principles} is based on symmetries and on the conservation equations that derive from them. 
Symmetries and their spontaneous breaking provide a natural route to classifying states of matter, formalized by Landau's theory of second-order phase transitions~\cite{landau2013statistical}. This is an example of effective field theory, valid around the critical temperature at which the phase transition occurs, where the relevant degrees of freedom are only the order parameter and its fluctuations. 

Hydrodynamics and its extensions to non-liquid states of matter (e.g. elasticity theory) \cite{chaikin2000principles} constitute another class of effective field theories, which describe the long-distance, late-time dynamics of the system. Microscopic degrees of freedom are integrated out in this limit, and are reorganized into fast and slow degrees of freedom. Fast degrees of freedom equilibrate on time and length scales short compared to the local equilibration scales, which are typically set by the temperature of the system. Slow degrees of freedom are protected by symmetries and need to be retained in the effective field theory -- they are the conserved densities of the system, such as energy, charge or momentum. Their evolution is described by conservation laws descending from the symmetries previously mentioned. They cannot decay locally and are transported away on scales much larger than the local equilibration scales to other regions of the system by hydrodynamic modes, such as sound or diffusion.

More concretely, the equations of motion for the conserved densities $n^a$ take the form\footnote{To keep the notation light, we are not including spatial indices for the moment, see section \ref{section:hydrodyn} for details.} 
\begin{equation}
\label{hydroeomsint}
    \dot n^a(t,x)+\nabla\cdot j^{a}(t,x)=0\,.
\end{equation}
Upper dots stand for time derivatives, $\dot{}\equiv\partial/\partial t$. For a fluid with a conserved $U(1)$ charge, the $n^a$'s are the set of energy $\varepsilon$, momentum $\pi^i$ and charge $n$ densities. 
The spatial currents $j^a=\{j_\varepsilon^i,\tau^{ij},j^i\}$ are generally not slow operators.\footnote{In special cases they can be. For example, in a system invariant under Galilean boosts, the corresponding Ward identity gives an operator equation between the charge current and momentum operators, so that the charge current is a slow operator in this case. In a relativistic system, the Lorentz boosts Ward identity equates the momentum and energy current operators, so the energy current becomes a slow operator as well.} They decay locally in the thermal bath of conserved densities, and therefore their 
expectation values over hydrodynamic timescales are tied to their overlap with the conserved densities via local expansions in terms of the densities and external sources:\footnote{Throughout this {\em Colloquium}, we work in a hydrodynamic frame where the time components of the conserved currents match the microscopic conserved densities, and dissipative corrections only enter with spatial gradients. Any time derivative correction can be traded for spatial derivatives by using the equations of motion and constitutive relations at lower order in derivatives.}
\begin{equation}
\label{jconstrelgen}
    \langle j_a\rangle=\alpha_{ab}^{(0)} \langle n^b\rangle+ \alpha_{ab}^{(1)} \nabla \langle n^b\rangle+\cdots
\,\end{equation}
The angular brackets denote a thermal average.
The $\alpha_{ab}^{(0),(1)}$ are transport coefficient matrices, order by order in the gradient expansion, with dots denoting higher-order terms.
Which of these coefficients are nonzero depends on the details of the system and the symmetry breaking pattern. The underlying reason why such expansions are possible is related to the central assumption to hydrodynamics: all microscopic, high energy modes relax on short scales of the order the thermalization time/length, and can be integrated out. At longer scales, only hydrodynamic fields are retained and are the sole source of non-analyticities in the retarded Green's functions. In other words, in the hydrodynamic regime, the retarded Green's functions only contain the gapless hydrodynamic poles.

In this \textit{Colloquium}, we will limit ourselves to expansions to first order in gradients. We will also ignore the effects of fluctuations, \cite{DESCHEPPER19741,Forster1977}, which generally spoil the analyticity of retarded Green's functions and of the dispersion relations of the hydrodynamic modes beyond first order in gradient terms. In gauge-gravity duality, these fluctuations are suppressed by the $N\to+\infty$ limit, \cite{Kovtun:2003vj}.

Inserting \eqref{jconstrelgen} in \eqref{hydroeomsint}, these become evolution equations for the vevs of the conserved densities, which can now be solved. Taking a spatial Fourier transform and dropping angular brackets for convenience, we obtain a set of dynamical equations given by
\begin{equation}
\label{hydroeomM}
    \dot n_a(t,q)+M_{ab}(q)\cdot n^b(t,q)=0\,.
\end{equation}
By construction, the matrix $M_{ab}(q)=M_1 q+M_2 q^2+\cdots$ has a local expansion in powers of the wave-vector $q$, with each term suppressed by the cutoff length of hydrodynamics $\ell_{th}$.

We would now like to compute the retarded Green's functions of the system. As usual, this implies turning on a time-dependent deformation of the Hamiltonian 
\begin{equation}
    H_o\mapsto H(t)=H_o-\int d^dx\, n^a(t,x)\delta\mu_{e,a}(t,x)
\end{equation}
(with $d$ the number of spatial dimensions), upon which the equations of motion become \cite{KADANOFF1963419,chaikin2000principles}
\begin{equation}
\label{hydroeomwithsources}
    \dot n^a(t,q)+M^a_b(q)\cdot \left(n^b(t,q)-\chi^b_{c}\,\delta\mu_{e}^c\right)=0\,.
\end{equation}
Here $\chi$ is the matrix of static susceptibilities, obtained by functional differentiation of the equilibrium free energy
\begin{equation}
\label{staticsusceptmatrixdef}
    \chi_{ab}(x-x')=-\frac{\delta^{2}W[\mu_e]}{\delta\mu_{e}^a(x)\delta\mu_{e}^b(x')}\,,
\end{equation}
where $W=-T\log \textrm{Tr} e^{-\beta H}$. This matrix encodes the linear response of the system to static perturbations $\delta\mu_{e}(x)$. In the static limit, from \eqref{hydroeomwithsources} $n^a=\chi^a_{b}\,\delta\mu_{e}^b$ and so $\chi_{ab}$ is simply the matrix of thermodynamic derivatives. It should be positive definite in order for the system to be locally thermodynamically stable.

Taking a Laplace transform of \eqref{hydroeomwithsources} (see \cite{Kovtun:2012rj} for more details) leads to the retarded Green's functions
\begin{equation}
    G^R_{ab}(\omega,q)\equiv\frac{\delta n^a(\omega,q)}{\delta\mu_{e}^b(\omega,q)}=-\left(i\omega-M\right)^{-1}\cdot M\cdot \chi
\end{equation}
where $\omega$ is the frequency.
The hydrodynamic poles of the system are found by solving the equation $\det(-i\omega+M)=0$. As a point of reference, in the case of a single conserved $U(1)$, the constitutive relation for the spatial current compatible with invariance under parity and time reversal and with external sources turned on is
\begin{equation}
    j^i=-D_n\left(\nabla^i n-\chi_{nn}\nabla^i\delta\mu_{e}\right)+\cdots,\quad i=1\ldots d\,,
\end{equation}
leading to a quadratically dispersing, diffusive mode $\omega=-iD_n q^2+\cdots$. The diffusivity can be measured by the following Kubo formula
\begin{equation}
\label{KuboChargeDiffusion}
    D_n=\frac1{\chi_{nn}}\lim_{\omega\to0}\lim_{q\to0}\frac\omega{q^2}\textrm{Im}G^R_{nn}(\omega,q)\,.
\end{equation}

Instead, in the longitudinal sector, a neutral fluid would have two linearly dispersing sound modes $\omega=\pm c_s q-i  \frac{\Gamma}{2} q^2$, where the longitudinal sound velocity is determined by the various static susceptibilities and the sound attenuation $\Gamma$ by first-order in gradients dissipative corrections to the constitutive relation of the energy current and stress tensor, see e.g. \cite{chaikin2000principles}.

Hydrodynamics only gives access to gapless poles with a vanishing dispersion relation at zero wavevector $\omega(q=0)=0$, the {hydrodynamic modes}. Non-hydrodynamic, gapped modes of the system cannot reliably be included in the hydrodynamics in general, except in certain special circumstances, for instance when the gap is generated by breaking weakly one of the symmetries of the system \cite{Davison:2014lua,Grozdanov:2018fic}. One of the goals of this \textit{Colloquium} is to explain how to incorporate such weakly-gapped degrees of freedom in the low energy effective field theory. Generic gapped modes which do not fall in the previous category typically signal the breakdown of the effective field theory description \cite{Withers:2018srf,Grozdanov:2019kge} and can only be accounted for by supplementing hydrodynamics with a microscopic completion.\footnote{Remarkably, when the gradient expansion can be systematically computed in a microscopically complete framework, the dispersion relation of gapped modes can be obtained by resumming the hydrodynamic series, \cite{Withers:2018srf}.}

%%%%%%%%%%%%%%%%%%%%%%%%%%%%
\section{Holographic methods}
\label{sec2}
The hydrodynamic construction outlined in the previous section can be systematically carried out order by order in the gradient expansion. 
The procedure quickly becomes intractable analytically due to the proliferation of terms to be considered, \cite{Grozdanov:2015kqa}. The equation of state and each transport coefficient needs to be measured experimentally or computed in a microscopic model.

Most microscopic models, nevertheless, face serious difficulties whenever the system under investigation is either strongly interacting, made of a large number of constituents, placed at finite chemical potential, finite temperature or when its real time dynamics is considered. Under these circumstances, the AdS-CFT correspondence\footnote{The acronyms stand for Anti de Sitter spacetime \cite{gibbons2000anti} and conformal field theory \cite{ginsparg1988applied} -- the two endpoints of the original strings-inspired holographic duality. In this \textit{Colloquium}, we will ignore more general situations in which the UV fixed point of the dual field theory is not a Lorentz invariant conformal field theory.} provides a self-consistent framework to attack these problems and guide new interdisciplinary explorations. Holography posits a duality between a large class of quantum field theories with gauge group of dimension $N$ and higher-dimensional gravitational theories (for details we refer to a number of reviews and textbooks now available in the literature \cite{Aharony:1999ti,Nastase:2007kj,natsuume2015ads,ammon_erdmenger_2015,zaanen2015holographic,Baggioli:2019rrs}). The duality was originally discovered in the context of string theory \cite{Maldacena:1997re,witten1998anti,Gubser:1998bc}, which provides a precise formulation of the conjecture, between a supersymmetric gauge theory ($\mathcal N=4$ super Yang-Mills with gauge group $SU(N)$) and a string theory (type IIB string theory on AdS$_5\times S^5$), now widely accepted as proven. The simplifying limit of classical gravity without extended objects corresponds to considering a dual field theory in the regime of strong coupling and in the large $N$ limit\footnote{Here, $N$ is the rank of the dual gauge field theory \cite{tHooft:2002ufq}. In the absence of a precisely identified dual field theory, this limit has to be understood as a large number of degrees of freedom. See \cite{zaanen2015holographic} for a discussion on this point.} and is known as the {bottom-up} approach. Bottom-up holographic methods have been applied in several directions such as quantum chromodynamics (QCD) and heavy ion collisions \cite{Casalderrey-Solana:2011dxg,Berges:2020fwq}, condensed matter many-body systems and quantum information \cite{zaanen2015holographic,Hartnoll:2016apf,rangamani2017holographic,Liu:2018crr,Liu:2020rrn}. 

From a formal point of view, the duality is built on the identification of the field theory generating functional $W$ with the gravitational on-shell path integral. The field theory operators  and sources are given by the specific coefficients of the asymptotic expansion of dynamical fields living in the curved, higher-dimensional, {bulk} spacetime. Thermal, finite density states in the dual field theory are captured by gravitational charged black hole solutions in the bulk, with the field theory temperature given by the Hawking temperature at the event horizon and the chemical potential by the boundary value of the bulk gauge field. From this gravitational background, all thermodynamic quantities can be computed, as well as the static susceptibilities. Linear perturbations of the gravitational solution together with opportune boundary conditions for the bulk fields \cite{Son:2002sd} yield the real-time, space-dependent retarded Green's functions, the poles of which are given by the quasi-normal modes of the black hole solution. This linear analysis also gives access to all linear transport coefficients through the appropriate Kubo formulas. This way, one can obtain the dispersion relations of the low-energy excitations in the dual field theory as well as those of the non-hydrodynamic modes of the system, going far beyond the hydrodynamic regime \cite{Kovtun:2005ev,Berti:2009kk}. Holographic results have been successfully matched to the predictions of charged, relativistic linearized hydrodynamics \cite{Policastro:2002se,Policastro:2002tn,Baier:2007ix,Erdmenger:2008rm,Banerjee:2008th}.\footnote{The full nonlinear structure of the hydrodynamic theory can be also derived from the gravitational equations using the {fluid-gravity correspondence} \cite{Rangamani:2009xk}.} Besides providing a concrete test bed for hydrodynamics, holography is a microscopically complete framework which allows to compute all transport coefficients, from which important lessons on strongly-coupled dynamics can be extracted. A case in point is the famous viscosity-entropy-ratio bound \cite{Policastro:2001yc,Cremonini:2011iq}.

As in the case of hydrodynamics, and more in general effective field theories, bottom-up holography is based and built on symmetries as guiding principles. Using the well-established holographic dictionary, local symmetries in the bulk are mapped into global symmetries of the boundary field theory. Any combination of explicit or spontaneous symmetry breaking can be considered. Explicit breaking corresponds to the presence of a source in the dual field theory which appears in Ward identities and spoils conservation equations; spontaneous breaking, on the contrary, is characterized by the appearance of a non-trivial vacuum expectation value for an operator -- the condensate -- which breaks (a subset of) the symmetries of the action from which it is derived \cite{Beekman:2019pmi}; finally, the pseudo-spontaneous regime appears when a small (to be quantified more precisely later) source is added on top of a purely spontaneous state \cite{PhysRevLett.29.1698,Burgess:1998ku}. From the bulk point of view, this distinction is encoded in the asymptotic behaviour of the field responsible for the symmetry breaking close to the boundary of the AdS spacetime \cite{Skenderis:2002wp}. The corresponding boundary Ward identities can be computed directly from the bulk as well (see for example \cite{Argurio:2015wgr} for the simplest case of a global $U(1)$ symmetry).

The holographic description of broken-symmetry, strongly-coupled phases of matter with an eye towards condensed matter was initiated in \cite{Gubser:2008px,Hartnoll:2008vx,Hartnoll:2008kx} by considering the spontaneous breaking of a global $U(1)$ symmetry -- a superfluid state.  Superfluid hydrodynamics correctly predicts the low energy dynamics of holographic superfluids, \cite{Herzog:2008he,Sonner:2010yx,Herzog:2011ec,Bhattacharya:2011eea,Bhattacharya:2011tra,Arean:2020nfa}.

Holographic lattices breaking translations explicitly were constructed numerically a few years later in \cite{Horowitz:2012ky,Horowitz:2012gs}.\footnote{See \cite{Chesler:2013qla,Donos:2014yya,Rangamani:2015hka,Langley:2015exa} for further numerical constructions of holographic lattices.} While the significance of explicit translation symmetry breaking was recognized early on in the holographic community \cite{Hartnoll:2007ih,Hartnoll:2008hs}, full holographic realizations had to tackle with the technical challenge related to solving inhomogeneous space-dependent Einstein's equations \cite{Dias:2015nua,Andrade:2017jmt,Krikun:2018ufr}. A change of paradigm happened with the discovery of the so-called {homogeneous models}, holographic setups in which translations are broken but the background metric and the dual stress tensor remain homogeneous (independent of the spatial coordinates). This property is due to existence of specific global structures which mix with spacetime symmetries, leading to a tremendous simplification in the computations of physical observables. The homogeneous models fall into different classes: (I) de Rham-Gabadadze-Tolley (dRGT) massive gravity theory \cite{Vegh:2013sk}\footnote{dRGT corresponds to a precise fine-tuned choice of more general Lorentz-violating massive gravity theories built in terms of St\"uckelberg fields \cite{Dubovsky:2004sg}. This fine tuning is not necessary around Lorentz-violating vacua. In this sense, the axion models in item (II) are not different from the dRGT model, which ultimately represents only an infinitesimal sub-class of them. See \cite{Alberte:2015isw} for a more detailed discussion of this point in the context of holographic models.}, (II) `axion' models \cite{Andrade:2013gsa,Taylor:2014tka, Gouteraux:2014hca,Donos:2014uba,Baggioli:2014roa,Baggioli:2021ejg} (see \cite{Baggioli:2021xuv} for a review), (III) Q-lattices \cite{Donos:2013eha}, (IV) higher-forms models \cite{Grozdanov:2018ewh}\footnote{Before being analyzed from a holographic point of view, static black hole solutions including matter in the form of free scalar and p-form fields were constructed in \cite{Bardoux:2012aw}.}, (V) helical lattices \cite{Nakamura:2009tf,Iizuka:2012iv,Donos:2012js,Donos:2014oha} and (VI) "solidon" models \cite{Esposito:2017qpj}. Irrespective of the specific holographic model employed, in the regime of weak explicit breaking, the low energy dynamics match well with the field theory expectations for a metallic phase with slowly-relaxing momentum, \cite{Hartnoll:2007ih,Hartnoll:2008hs,Davison:2013jba,Lucas:2014zea,Davison:2014lua,Lucas:2015pxa}. A common feature of all holographic model is that the graviton acquires a mass, \cite{Vegh:2013sk,Blake:2013owa,Alberte:2015isw}.\footnote{See \cite{Zaanen:2021zqs} for a discussion about the connections between elasticity theory and (massive) gravity.}

\begin{figure}
    \centering
    \includegraphics[width=\linewidth]{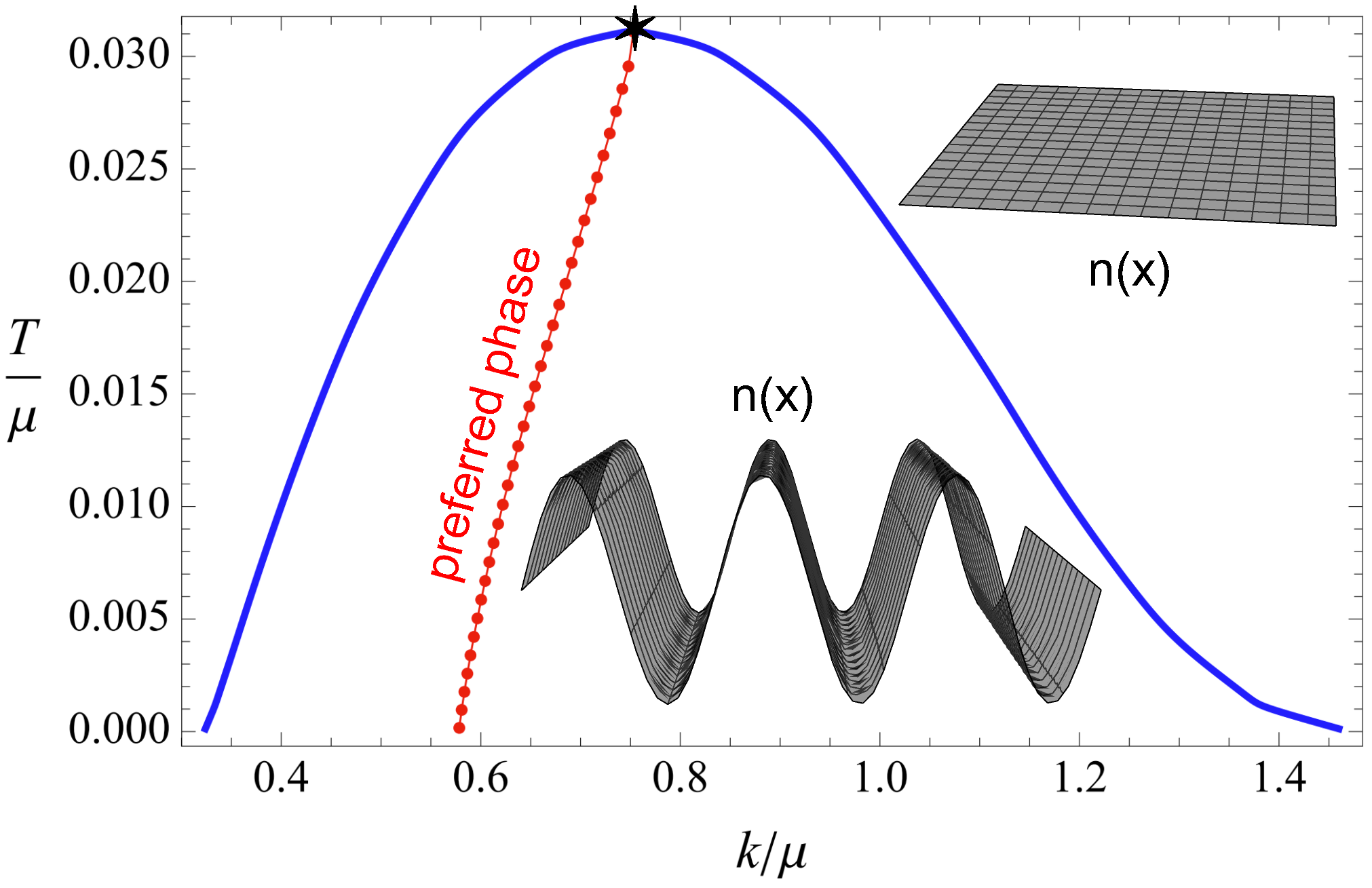}
    \caption{The typical curve describing the onset of the instability towards a holographic phase breaking translations spontaneously. The insets show the spatial profile of the charge density $n$ in the broken and normal phases. Figure adapted with permission from \cite{Cai:2017qdz}.}
    \label{fig:bell}
\end{figure}
Closely following the global $U(1)$ case, instabilities towards holographic spatially modulated phases breaking translations spontaneously were investigated as well, \cite{Nakamura:2009tf,Donos:2011bh,Donos:2011qt,Cremonini:2012ir,Alsup:2013kda}. In these studies, one looks for a spatially modulated, normalizable bulk mode (i.e., without a source at the boundary) in the translation-invariant, homogeneous bulk geometry. The outcome of this analysis is an instability curve displaying the maximum temperature at which the mode can be found vs the wavevector of the modulation (see figure \ref{fig:bell}), i.e., the onset of the instability. The apex of this curve gives the thermodynamically preferred wavevector with the highest critical temperature, below which a fully backreacted, spatially modulated phase breaking translations spontaneously can be expected to be found. As such, when the preferred trajectory within the instability curve is followed (red bullets in figure \ref{fig:bell}), these phases are true global minima of the thermodynamic free energy, \cite{Donos:2013cka}. The breaking of parity and time reversal through Chern-Simons couplings in the bulk and/or an external magnetic field played an important role originally to generate the instabilities, but is not always necessary, \cite{Donos:2013gda}. The original works focused on inhomogeneous instabilities, but helical phases proved easier to construct at first, \cite{Donos:2011ff,Donos:2012gg,Donos:2012wi}. Backreacted inhomogeneous phases spatially modulated along one direction were constructed in \cite{Rozali:2012es,Donos:2013wia, Withers:2013loa}, bearing on the expertise developed to construct explicit holographic lattices. Generalizations to two-dimensional, checkerboard or triangular patterns are found in \cite{Withers:2014sja,Donos:2015eew}, with the triangular lattice providing the minimum free energy state. Remarkably, these phases all include circulating current loops together with spontaneous parity breaking, which is reminiscent of the loop current order proposed to underlie the pseudogap phase of underdoped cuprate high $T_c$ superconductors, \cite{PhysRevLett.83.3538}. This phenomenology is a direct consequence of the bulk Chern-Simons term.

Probe branes constructions can also display spatially modulated instabilities \cite{Jokela:2012se,Jokela:2014dba,Jokela:2016xuy}. Being top-down models descending from specific string theory realizations, they have the advantage of offering a more precise field theory interpretation. On the other hand, it is not clear how one should interpret the spontaneous spatial modulation of charge and current densities, since in these setups momentum and temperature fluctuations are frozen.

Phases in which the breaking of translations and of a global $U(1)$ are intertwined are of interest to model pair density wave phases \cite{Fradkin:2015}, which are thought to play an important role in the phase diagram of underdoped cuprate high $T_c$ superconductors. They have been argued to be the mother phase from which daughter charge density wave and superconducting phases emerge. Holographic realization of these phases are found in \cite{Cremonini:2016rbd,Cremonini:2017usb,Cai:2017qdz}. These constructions rely on a combination of bulk Chern-Simons terms and the introduction of St\"uckelberg scalars, which naturally give rise to pair density wave phases where the condensate is spatially modulated with a zero average and periodicity which is twice that of the charge density wave.\footnote{See also \cite{Baggioli:2022aft} for a homogeneous model realizing the concomitant spontaneous breaking of translations together with a global $U(1)$ symmetry -- a supersolid phase.}

Natural next steps were to combine all of these strands together, by considering holographic phases breaking translations pseudo-spontaneously and how their low energy dynamics matched to field theory expectations. The purpose of this \textit{Colloquium} is to summarize the progress in these directions over the last few years. These developments came about from the intersection between various pieces of work: the incorporation of the physics of explicit symmetry breaking in the hydrodynamics of phases breaking translations spontaneously \cite{Delacretaz:2016ivq,Delacretaz:2017zxd,Delacretaz:2021qqu,Armas:2021vku}; the construction of simpler homogeneous holographic models for (pseudo-)spontaneously breaking translations \cite{Amoretti:2016bxs,Alberte:2017cch,Andrade:2017cnc,Alberte:2017oqx,Amoretti:2017frz,Amoretti:2018tzw,Donos:2019tmo,Ammon:2019apj,Ammon:2019wci,Donos:2019hpp}, which provided a far more tractable platform to compare with hydrodynamic predictions; a thorough analysis of how background strain and external sources enter in the hydrodynamic theory, \cite{Armas:2019sbe, Armas:2020bmo}, and subsequent comparison with holographic constructions \cite{Ammon:2020xyv}.

%%%%%%%%%%%%%%%%%%%%%%%%%%%%%%%%%
\section{Hydrodynamics of phases with broken translations}
\label{sec11}

Continuous, global symmetries can be spontaneously broken in the ground state (see \cite{Beekman:2019pmi} for a recent introduction). Formally, this means that the ground state is invariant under a smaller set of symmetries than the Hamiltonian of the system. A corresponding number of gapless modes, the Goldstone bosons, appear in the spectrum. In the simplest case of an internal symmetry and when the broken generators commute, the number of Goldstones is given by the number of broken generators. In other cases, such as that of spacetime symmetries \cite{PhysRevLett.88.101602}, the counting rule is more complicated, see \cite{Watanabe:2019xul} for a review.

Hydrodynamics can be advantageously extended to systems with spontaneously broken symmetries, like superfluid Helium or crystalline solids, \cite{chaikin2000principles}. The set of slow degrees of freedom is enlarged to include the Goldstone modes, the dynamical evolution of which is described by so-called Josephson equations (historically, the Josephson equation describes the phase difference in a superconductor in the presence of an external voltage). 

In this \textit{Colloquium}, we focus on the case of broken translations, such as crystalline solids, charge density waves and Wigner crystals. We incorporate the effects of background strain, which proves important to match to existing holographic studies. Moreover, strain/pressure is also a common experimental tool in the investigation of broken translation phases in strongly-correlated materials and has a strong effect on the onset of the charge density wave and superconducting phases, \cite{Mackenzie:2014,LeTacon:2018}.  We also do not assume any particular boost symmetry.\footnote{In this review, we will not consider equilibrium states with a background fluid velocity and so will only work at linear order in the fluid velocity. See \cite{deBoer:2017ing,deBoer:2017abi,Novak:2019wqg,Poovuttikul:2019ckt,deBoer:2020xlc,Armas:2020mpr} for fully nonlinear treatments of fluid hydrodynamics without boosts.}

\subsection{Thermodynamics}

For simplicity, we assume the system is two-dimensional ($d=2$), isotropic and that translations are spontaneously broken in all spatial directions -- extensions to anisotropic or higher-dimensional crystals are conceptually straightforward but technically tedious, due to more complicated tensor structures and a larger number of transport coefficients. We will not consider the coupling to background sources, which can be realized along the lines of \cite{Armas:2019sbe,Armas:2020bmo}.

Since spatial translations are spontaneously broken in all directions, we expect as many Goldstone modes as there are broken translations.\footnote{Rotations are also spontaneously broken, but do not have independent Goldstones, {per} the Goldstone counting theorems for broken spacetime symmetries \cite{PhysRevLett.88.101602}. The underlying reason is that translations and rotations are not independent local transformations.}  The Goldstones of broken translations are often called phonons and are related to the displacements of the underlying crystal structure, which we denote by $u^i$. They transform non-linearly under spatial translations $x^i\to x^i+a^i$ as
\begin{equation}
\label{ushift}
    u^i\to u^i+a^i\,.
\end{equation}
The free energy of the system must be invariant under these shifts, and so the displacements can appear therein only with derivatives. From a Lagrangian perspective,\footnote{This assumption is important only at nonlinear level since all different definitions agree with each other at the linear one \cite{ZAMM:ZAMM19850650903}.} we define the non-linear, Lagrange strain tensor $u_{ij}$ as
\begin{equation}
    dx'^2-dx^2\equiv 2\,u_{ij}dx^i dx^j\,.
\end{equation}
Here $x'(x)=x+u(x)$ is the new location of the point originally at $x$ after deforming by a small displacement $u$. 
Writing that $dx'=x'(x+dx)-x'(x)$, the nonlinear strain is then given in terms of the displacements by \cite{chaikin2000principles}
\begin{equation}\label{straindef}
u_{ij}=\nabla_{(i}u_{j)}+\frac12\nabla_i u_k\nabla_j u^k\,.
\end{equation}
The free energy only depends on $u_{ij}$. Lower-case latin indices $i,j,\dots$ are raised an lowered with the Kronecker delta $\delta_{ij}$ and run over spatial dimensions.
 
The elastic part of the equilibrium free energy density of the system is then, by isotropy,
\begin{equation}
\label{elasticfreeenergy}
f_{el}=\frac{B_o(X,Y)}2\,X^2+G_o(X,Y)\,Y
\end{equation}
where we have defined $X\equiv u_i^i$, $Y\equiv u_{ij}u^{ij}-\frac{1}{2} \left(u_i^i\right)$ and suppressed the dependence on temperature and chemical potential for now.\footnote{One could equivalently define a generic function $f_{el}(X,Y)$. Our parametrization makes the linear limit $X,Y \rightarrow 0$ and the limit of zero background strain clearer. In $d>2$, this function would depend on $d$ independent scalars, \cite{Nicolis:2013lma,Esposito:2017qpj}.} The first parameter $X$ corresponds to a purely volumetric deformation, $\Delta V/V = X$, while the second, $Y$, to a deviatoric deformation which modifies the shape of the material but not its total volume. Accordingly, the coefficients $B_o,G_o$ are the bare nonlinear bulk and shear moduli, respectively. If translations are broken in one dimension only, there is only a bulk modulus. Both quantities are nonlinear functions of the deformation parameters $X,Y$ and, in what follows, of temperature and chemical potential as well.

Since we wish to compute the linear response of the system to external perturbations, the first step is to determine the static susceptibilities. To this end, we expand the static free energy to quadratic order in fluctuations using\footnote{This maps to the formulation of \cite{Armas:2019sbe} as follows: $u^i\leftrightarrow \Phi^I-x^I$, $u^{ij}\leftrightarrow h^{IJ}-\delta^{IJ}$, $m=\frac1\alpha-1$.} 
 \begin{equation}
\label{linearizeddisp}
u_i = m\, x^i+(1+m)\delta\phi_i(x)\,,\quad i=\{x,y\}\,,
\end{equation}
where $m$ is a real parameter, and we have assumed for simplicity that the linear perturbations $\delta\phi_i$ depend only on one spatial dimension. 

The first term $m x^i$ in Eq.\eqref{linearizeddisp} can be thought as the additional displacement from the would-be static equilibrium configuration $m=0$ to the actual configuration with isotropic background strain $m\neq 0$:\footnote{Anisotropic strains can easily be considered by allowing for additional off-diagonal terms of the type $u_i \sim x_j$ (see e.g. \cite{Baggioli:2020qdg}).} \begin{equation}
X=u_l^l=u_o+\mathcal O(\nabla)\,,\quad u_o\equiv m(m+2)\,.  
\end{equation}
For this state, $Y=\mathcal O(\nabla^2)$. In this configuration, the equilibrium free energy density $f_{el}$, and the bare elastic moduli $B_o$, $G_o$ are functions of $u_o$. As we are about to see, the $\delta\phi_i$ are the Goldstones modes of the system around the configuration with background strain $u_o$.

Plugging in the expansion \eqref{linearizeddisp} in the elastic free energy \eqref{elasticfreeenergy} and expanding to quadratic order in fluctuations, we obtain
\begin{equation}
\label{elasticfequad}
f_{el}=\frac{u_o^2}2B_o-p_{el}\lambda_\parallel+\frac{G}2\left(\lambda_\perp\right)^2+\frac12\left(B+G\right)\left(\lambda_\parallel\right)^2
\end{equation}
where it is convenient to define the longitudinal and transverse Goldstones $\lambda_\parallel=\nabla\cdot\delta\phi$, $\lambda_\perp=\nabla\times\delta\phi$. In the presence of nonzero background strain $u_o$, the free energy has a term linear in $\lambda_\parallel$, which defines the background elastic pressure $p_{el}=-(1+u_o)\partial_{u_o}(u_o^2 B_o)/2$. It is manifest that the configuration with zero strain minimizes the free energy, so that states with finite background strain must be sourced by non-trivial boundary conditions. The bulk and shear moduli also pick up new contributions
\begin{equation}
\label{renormBG}
\begin{split}
B\,\equiv&\,\frac12(1+u_o)^2\partial_{u_o}^2\left(u_o^2 B_o\right)\,,\\
G\,\equiv&\,(1+u_o)^2G_o+\frac{u_o^2}{2}(1+u_o)^2\partial_Y B_o-p_{el}\,.
\end{split}
\end{equation}
In the limit of zero strain, we recover $B(u_o=0)=B_o$ and $G(u_o=0)=G_o$.

The free energy in Eq.\eqref{elasticfequad} now displays a linear term in $\delta\phi_x$ with coefficient $p_{el}$.\footnote{This term also appears in the relativistic treatment of \cite{Armas:2019sbe}. Here we generate it by expanding around the state with background strain in Eq.\eqref{linearizeddisp}. In \cite{Armas:2019sbe}, $p_{el}$ is introduced directly as a force contribution to the free energy. There, the reference contribution is not assumed to minimize the free energy. After matching conventions, both approaches give the same results.} This linear term implies that the system is not in mechanical equilibrium whenever  $m$ is nonzero: a background strain is applied to the system, through non-trivial boundary conditions and the resulting sum of external forces does not vanish. Instead, in mechanical equilibrium, there is no background displacement, $m=0$, and  $p_{el}=0$. The standard treatment of elasticity theory (see e.g. \cite{chaikin2000principles}) assumes the reference configuration, the choice of which is arbitrary, to correspond with the state of mechanical equilibrium. Nevertheless, this new term might be relevant for many experiments which study phases with broken translations under the application of pressure. It appears as well in the viscoelastic description of pre-strained materials \cite{biot1965internal,doi:10.1063/1.1712807,doi:10.1063/1.1710417,berjamin2021acoustoelastic,destrade2007waves,destrade2009small,doi:10.1121/1.1911738} and it has been recently considered in the relativistic viscoelastic framework of \cite{Armas:2019sbe} under the name of crystal or lattice pressure.\\

Turning on background external sources $f\mapsto f-s_{\parallel}\lambda_\parallel-s_{\perp}\lambda_\perp$, we obtain the Goldstone static susceptibilities after integrating out the Goldstones:\footnote{This leads to $s_{\perp}=G \lambda_{\perp}$, $s_{\parallel}=(G+B) \lambda_{\parallel}$.}
\begin{equation}
\label{GoldstoneStaticSusceptibilities}
\begin{split}
    &\chi_{\lambda_\parallel\lambda_\parallel}\equiv-\frac{\partial^{(2)}f_{el}}{\partial s_\parallel{}^2}=\frac{1}{B+G}\,,\\
    & \chi_{\lambda_\perp\lambda_\perp}\equiv-\frac{\partial^{(2)}f_{el}}{\partial s_\perp{}^2}=\frac{1}{G}\,.
    \end{split}
\end{equation}
$B+G$ and $G$ should both be positive definite in order for the phase to be locally thermodynamically stable, which follows from the usual requirement that the determinant of the Hessian of the free energy be positive definite. We will see later on that this ensures that sound modes have a positive velocity squared.
Through \eqref{renormBG}, we observe that both the bulk and shear moduli have a non-trivial dependence on background strain. Varying the background strain may lead to thermodynamic instabilities, signaled by divergences in the static susceptibilities \eqref{GoldstoneStaticSusceptibilities} when $G$ or $B+G$ change sign. Determining whether these instabilities are actually present requires to know their functional dependence on $u_o$ and is beyond the effective field theory approach.

By their properties under parity transformations $x\mapsto-x$, we also expect the longitudinal phonon $\lambda_\parallel$ to couple to entropy and charge. To this end, we include temperature and chemical potential dependence in the bare moduli $B_o(u_i^i,T,\mu)$ and $G_o(u_i^i,T,\mu)$ in \eqref{elasticfreeenergy}. Linearizing around \eqref{linearizeddisp} together with $\left\{T,\mu\right\}=\left\{T_o,\mu_o\right\}+\left\{\delta T,\delta\mu\right\}$ allows to identify the off-diagonal susceptibilities
\begin{equation}
\begin{split}
&\chi_{n\lambda_\parallel}\equiv-\frac{\partial^{(2)}f_{el}}{\partial s_\parallel\partial\mu}=\frac{\partial_{\mu}p_{el}}{B+G}\,,\\
 &\chi_{s\lambda_\parallel}\equiv-\frac{\partial^{(2)}f_{el}}{\partial s_\parallel\partial T}=\frac{\partial_{T}p_{el}}{B+G}\,.
 \end{split}
\end{equation}
They are nonzero even in the absence of background strain and correspond physically to the chemical and thermal expansion of the system under strain.

The full longitudinal static susceptibility matrix reads:
\begin{equation}
\label{longstatsusc}
\chi_{o,\parallel}=\left(
    \begin{array}{cccc}
         \chi_{nn}&\chi_{n\varepsilon}&0&\chi_{n\lambda_\parallel} \\
         \chi_{n\varepsilon}&\chi_{\varepsilon\varepsilon}&0&\chi_{\varepsilon\lambda_\parallel}\\
         0&0&\chi_{\pi\pi}&0\\
         \chi_{n\lambda_\parallel} &\chi_{\varepsilon\lambda_\parallel}&0
         &\chi_{\lambda_\parallel\lambda_\parallel} 
    \end{array}\right)\,.
\end{equation}
The equality of off-diagonal components follows from invariance under PT symmetry. 

In the transverse sector, the susceptibility matrix $\chi_{o,\perp}$ is diagonal with the two nonzero elements $\chi_{\lambda_\perp\lambda_\perp}$, given by \eqref{GoldstoneStaticSusceptibilities}, and $\chi_{\pi_\perp\pi_\perp}=\chi_{\pi\pi}$, by isotropy.

\subsection{Dynamics\label{section:hydrodyn}}

We are now ready to state the equations that govern the dynamics of the system in the hydrodynamic regime. Assuming rotation, translation and $U(1)$ symmetry, these are the conservation of energy, charge and momentum density
\begin{equation}\label{hydroeoms}
    \dot\varepsilon+\nabla\cdot j_\varepsilon=0\,,\quad \dot n+\nabla\cdot j=0\,,\quad \dot\pi^i+\nabla_j\tau^{ji}=0\,,
\end{equation}
together with the Josephson equation for the dynamic evolution of the Goldstones:
\begin{equation}
\label{JosRelIdealLambda}
    \frac{\mathrm d}{\mathrm{dt}}u^i=- v^i+\dots\,.
\end{equation}
Here, $v^i$ is the velocity field conjugate to momentum $\pi^i$, $\mathrm d/\mathrm{dt}\equiv\partial_t+v^i\nabla_i$ stands for the material derivative and the dots for dissipative corrections to this relation. 

%We can arrive at
We can derive the non-dissipative terms in \eqref{JosRelIdealLambda} in the following way. The Goldstone fields are canonically conjugate to the momentum density, i.e. the conserved charge that generates the broken symmetry:
\begin{equation}\label{commutatorgoldstone}
    i\,[\pi^i(x),u^j(x')]=-\delta^{(2)}(x-x')\left(\delta^{ij}+\nabla^i u^j\right)\,.
\end{equation}
Then, we deform the Hamiltonian by an external velocity source $H_o\mapsto H=H_o-\int d^2 x\, \pi_i v^i_{e}$ and use the Schr\"odinger equation to compute the time evolution of the displacement:
\begin{equation}
    \dot u^i=i\,[H,u^i]=v^i_{e}+v^j_{e}\nabla_j u^i\,.
\end{equation}
Since $u^i$ must be time-independent in thermodynamic equilibrium ($v^i=v_e^i$), this means that the Josephson relation must take the form
\begin{equation}
\label{JosRelIdealSources}
    \dot u^i=(v^j_e-v^j)\left(\delta _j^i+\nabla_j u^i\right)+\tilde u^i\,,
\end{equation}
in agreement with \eqref{hydroeomwithsources}. Taking a divergence or a curl of \eqref{JosRelIdealSources} with sources off leads to \eqref{JosRelIdealLambda}. We have allowed for a possible dissipative correction $\tilde u^i$.

In our thermodynamic ensemble, the first law of thermodynamics is
\begin{equation}
df=-s \,dT-n\, d\mu+h^{ij}d\left(\nabla_i u_j\right),
\end{equation}
where $h^{ij}\equiv\partial f/\partial(\nabla_i u_j)$ is
\begin{equation}
\begin{split}
h^{ij}=&\left(XB_o+\frac{X^2}2\partial_X B_o+Y\partial_X G_o\right) X^{ij}\\
&+\left(2G_o+Y\partial_Y G_o+\frac{X^2}{2}\partial_Y B_o\right) Y^{ij}\,,
\end{split}
\end{equation}
with 
\begin{equation}
\begin{split}
   X^{ij}&=\frac{\partial X}{\partial\nabla_i u_j}=\delta^{ij}+\nabla^i u^{j}\,,\\Y^{ij}&=\frac{\partial Y}{\partial \nabla_i u_j}=2\,(u^{ij}+u^{ik}\nabla_k u^j)-X X^{ij}\,.
   \end{split}
\end{equation}
Using that the entropy density must be conserved $\dot s+\nabla_i(sv^i+\tilde j_q^i/T)=0$ in the absence of dissipative (gradient) corrections, the ideal constitutive relations are found to be
\begin{align}\label{constrelideal}
j_\varepsilon^i=&(\varepsilon+p)\,v^i+h^{ij}v_j+h^{il}v^j\nabla_{j}u_l+\tilde j_\varepsilon^i\,,\\
 \tau^{ij}=&p\,\delta^{ij}+h^{ij}+h^{il}\nabla^{j}u^l+v^i\pi^j+\tilde\tau^{ij}\,,\\
 j^i=&n\,v^i+\tilde j^i\,.
\end{align}
Here $p$ is the thermodynamic pressure, which verifies $p=-f=-\varepsilon+s T+n\mu+v_k \pi^k$.
It is straightforward to verify that the stress tensor $\tau^{ji}$ is symmetric by substituting the expression for $h^{ij}$ in terms of $u^{ij}$ in \eqref{constrelideal}. $\tilde j_q^i$, $\tilde j_\varepsilon^i$, $\tilde \tau^{ji}$ and $\tilde j^i$ all stand for dissipative corrections which are at least first order in gradients. 

The form of dissipative corrections are determined by a well-known algorithm. We start by allowing all possible terms that are spatial derivatives of the fields (the conserved densities and the Goldstones) consistent with the symmetries -- for instance, we do not allow terms that violate parity. Then, we require that these terms do not lead to non-localities in the equations of motion. Finally, we check that the entropy current is positive definite, also imposing Onsager relations. The outcome of this procedure, which we detail in appendix \ref{app:entropyproduction}, leads to the following constitutive relations:
\begin{equation}
\label{constreldiss}
\begin{split}
&\tilde j^i=-\sigma_o^{ij}\nabla_j\mu-\alpha_o^{ij}\nabla_j T-\frac12\xi_\mu^{ij}\nabla^k h_{kj}\,,\\
&\frac{\tilde j_q^i}T=-\alpha_o^{ij}\nabla_j \mu-\frac{\bar\kappa_o^{ij}}{T}\nabla_j T-\frac12\xi_T^{ij}\nabla^k h_{kj}\,,\\
&\tilde\tau^{ij}=-\eta^{ijkl}\nabla_{(k}v_{l)}\,, \\
&\tilde u^i=\xi_{\mu}^{ij}\nabla_j\mu+\xi_T^{ij}\nabla_jT+\xi_h^{ij}\nabla^k h_{kj}\,,\\
&\tilde j_\varepsilon^i=\tilde j_q^i+\mu \tilde j^i-h_{ij}\tilde u^j+v_j\tilde\tau^{ij}\,.
\end{split}
\end{equation}
In the absence of background strain and to linear order in the fluid velocity, all the transport matrices would have a trivial index structure and depend on temperature and chemical potential only, e.g. $\sigma_o^{ij}=\sigma_{(o)}(T,\mu) \delta^{ij}$ or $\eta^{ijkl}\nabla_{(k}v_{l)}=-\eta\sigma^{ij}-\frac2d\zeta\partial_k v^k\delta^{ij}$, where we have defined the shear rate tensor $\sigma_{ij}=\nabla_{(i}v_{j)}-\frac2d\nabla_k v^k\delta_{ij}$.

In the presence of background strain, the strain tensor $u_{ij}$ provides an independent rank-2 tensor. This gives rise to new terms in the transport matrices, which all take the form $\sigma_o^{ij}=\sigma_{(o)}(T,\mu,u_o)\delta^{ij}+\sigma_{(u)}(T,\mu,u_o)u^{ij}$ in $d=2$\footnote{In $d>2$, additional tensor structures such as $u^i_{k}u^{kj}$ would enter. To map to the formulation of \cite{Armas:2019sbe}, higher-order terms in their strain $u^{IJ}$ need to be considered.} with some arbitrary dependence on $u_o$ (since $X=u_o$ and $Y=0$ when evaluated on the background \eqref{linearizeddisp}). There is more freedom in the viscosity rank-4 tensor, which takes the general form
\begin{equation}
\begin{split}
\eta_{ijkl}=&2\,\eta^{(0)}\left(\delta_{ik}\delta_{jl}-\frac12\delta_{ij}\delta_{kl}\right)+\zeta^{(0)}\delta_{ij}\delta_{kl}\\
&+2\,\eta^{(u)}\left(\delta_{ik}u_{jl}-\frac12\delta_{ij}u_{kl}-\frac12u_{ij}\delta_{kl}+\frac14 u_m^m\delta_{ij}\delta_{kl}\right)\\
&+2\left(\zeta^{(u)}+\bar\zeta^{(u)}\right)\delta_{ij}u_{\langle kl\rangle}+2\left(\zeta^{(u)}-\bar\zeta^{(u)}\right)u_{\langle ij\rangle}\delta_{kl}
\end{split}
\end{equation}
where angular brackets stand for the transverse, traceless part of the tensor.

\subsection{Linear response}
 
With the constitutive relations in hand, we can now investigate the linear response of the system about the equilibrium state (denoted with a $o$ subscript). Making use of  the underlying translation invariance of the system to decompose the linear perturbations in plane waves, we take $n^a=n^a_o+\delta n^a  e^{-i\omega t+i q x}$, $\mu^a=\mu_o^a+\delta\mu^a e^{-i\omega t+i q x}$, where the $n^a$ are the various conserved densities, and the $\mu^a$ their conjugate sources. We will not consider a background fluid velocity in this review.

We start with the transverse sector. In contrast to the fluid case discussed at the end of Section \ref{sec1} (see also \cite{Kovtun:2012rj}), the transverse Goldstone field mixes with transverse momentum to form a pair of sound modes propagating in opposite directions:
\begin{equation}
\label{TransverseSoundMode}
    \omega=\pm\, q\,\sqrt{\frac{G}{\chi_{\pi\pi}}}-\frac{i}2\left(\frac{\eta}{\chi_{\pi\pi}}+G\,\xi\right)q^2+O(q^3)\,.
\end{equation}
This is the celebrated shear sound mode of crystalline solids. Its velocity is real provided that the matrix of static susceptibility is positive definite, which implies $G>0$ and $\chi_{\pi\pi}>0$. The sound attenuation receives two contributions
\begin{equation}
\label{transpcoeffstransredef}
    \eta\equiv\eta^{(0)}+\frac{u_o}{2}\eta^{(u)}\,,\quad\xi\equiv\frac1{1+u_0}\left(\xi_h^{(0)}+\frac{u_0}2\xi_h^{(u)}\right)\,.
\end{equation}
Here, as in the longitudinal sector, we find that the effect of the extra terms in the constitutive relations due to background strain can be hidden away in a redefinition of the transport coefficients contributing to linear response. This is advantageous as this means there is no proliferation of transport coefficients. For instance, the shear Kubo formula that usually measures the shear viscosity for a fluid becomes
\begin{equation}
    \eta^{(0)}+\frac{u_o}{2}\eta^{(u)}\equiv\eta=-\lim_{\omega\to0}\frac{1}{\omega}\textrm{Im}G^R_{\tau^{xy}\tau^{xy}}(\omega,q=0)\label{thiseq}
\end{equation}
and it is the linear combination \eqref{transpcoeffstransredef} which appears, not the individual transport coefficients $\eta^{(0)}$ and $\eta^{(u)}$.

There is a similar Kubo formula for $\xi_\mu$:
\begin{equation}
\label{Kuboximu}
    \xi_\mu=-\lim_{\omega\to0}\frac1\omega\textrm{Im}G^R_{j\dot u}(\omega,q=0)
\end{equation}
and for $\xi$:
\begin{equation}
\label{Kuboxi}
    \xi=-\lim_{\omega\to0}\frac1\omega\textrm{Im}G^R_{\dot u\dot u}(\omega,q=0)
\end{equation}
which as we will see defines the Goldstone diffusivity.

In the longitudinal sector, there are four modes: two sound modes propagating in opposite directions and two diffusive modes. Their expressions are in general quite complicated, and so we report them only for a neutral, relativistic system (in which one of the diffusive mode disappears)\footnote{See the appendix \ref{app:entropyproduction} for details on how to take this limit. See also \cite{Armas:2020bmo} for the complete expressions at finite density.} 
\begin{equation}
\label{LongitudinalModes}
\begin{split}
    &\omega=\pm\, c_\parallel q-\frac{i}2 \Gamma_\parallel q^2\,,\quad\omega=-i D_\parallel q^2\,,\\
    &c_\parallel{}^2=\frac{B+G}{\chi_{\pi\pi}}+\frac{T^2\left(s_o-\chi_{s h_\parallel}\right)^2}{\chi_{\pi\pi}\chi_{\varepsilon\varepsilon}^{(\lambda_\parallel)}}\,,\\
    & \Gamma_\parallel =\frac{\eta+\zeta}{\chi_{\pi\pi}}+\frac{\xi\,\chi_{\pi\pi}}{c_\parallel{}^2}\left(c_\parallel{}^2-\frac{T\left(s_o-\chi_{s h_\parallel}\right)}{\chi_{\varepsilon\varepsilon}^{(\lambda_\parallel)}}\right)^2\,,\\
    &D_\parallel=\frac{(B+G)\,\chi_{\pi\pi}\,\xi}{c_\parallel{}^2\chi_{\varepsilon\varepsilon}^{(\lambda_\parallel)}}\,.
    \end{split}
\end{equation}
Here $\chi_{sh_\parallel}=\partial s/\partial\lambda_\parallel=-\partial p_{el}/\partial T_0$, and similarly $\chi_{\varepsilon\varepsilon}^{(\lambda_\parallel)}= c_v T_0$ (with $c_v$ the heat capacity), computed fixing $\lambda_\parallel$.
Similarly to the transverse sector, only certain linear combinations of transport coefficients appear (e.g. $\xi$ instead of both $\xi_{h}^{(0)}$ and $\xi_h^{(u)}$). After matching conventions, these expressions agree with \cite{Armas:2019sbe, Armas:2020bmo}. The modes can be worked out in full generality (absence of boost symmetry, finite density, nonzero background strain), but become rather complicated. The appearance of instabilities related to a change of sign of $B+G$ is manifest in the expression for $D_\parallel$, as the corresponding purely imaginary mode would then cross to the upper half complex frequency plane.

At very low temperatures, we expect the sound modes to be mostly carried by the Goldstone field and the longitudinal momentum density, while the diffusive mode corresponds to thermal diffusion. At temperatures close to the critical temperature, the sound modes are carried by thermal and momentum fluctuations, while the diffusive mode is predominantly carried by the Goldstone mode.\footnote{This could be verified explicitly by deriving the fluctuations eigenvectors of the hydrodynamics equations and also from the holographic computation of quasinormal modes, as done for a superfluid in \cite{Arean:2020nfa}.}

%%%%%%%%%%%%%%%%%%%%%%%%%%%%%%%%%%%%%%%
\subsection{Holography\label{section:HoloSp}}

Elastic properties in holographic models with broken translations have been investigated for some time \cite{Alberte:2015isw,Alberte:2016xja}. While hydrodynamics for Galilean-invariant phases with broken translations is an old subject \cite{PhysRevA.6.2401,chaikin2000principles,fleming1976hydrodynamics,RevModPhys.46.705}, it was revisited in \cite{Delacretaz:2017zxd}, which incorporated phase relaxation by defects and pinning by disorder in a hydrodynamic framework without assuming Galilean invariance. This led to a flurry of activity in the holographic community, intent on verifying the match between the holographic and hydrodynamic approaches, \cite{Alberte:2017oqx,Amoretti:2017axe,Amoretti:2019cef,Ammon:2019apj,Baggioli:2019abx,Armas:2019sbe,Ammon:2020xyv,Armas:2020bmo,Baggioli:2020edn}. Ultimately, this lead to a consistent hydrodynamic construction with nonzero background strain and with coupling to external sources, \cite{Armas:2019sbe,Armas:2020bmo}.

The presence of an isotropic background strain (equivalently, a background elastic pressure $p_{el}$) is a common feature of homogeneous holographic models based on massive gravity or Q-lattices,\footnote{The reason for this is clear: in the bulk, the fluctuations of the massless scalars are dual to the Goldstone modes, \cite{Donos:2019tmo,Amoretti:2019cef,Ammon:2019apj}, and their background Ansatz is chosen to be $\phi^i(r,x^i)=m x^i$ in order for homogeneity to be preserved. Their action is a bulk version of the EFT elastic free energy \eqref{elasticfreeenergy}, \cite{Nicolis:2013lma,Nicolis:2015sra}, and the bulk Ansatz can be directly compared to \eqref{linearizeddisp}, leading to a nonzero background strain. If the bulk parameter $m$ is set to zero, translations are no longer broken in any way.} which at an operational level can be directly observed by identifying an extra contribution $p_{el}$ to the (relativistic) momentum susceptibility, $\chi_{\pi\pi}=\varepsilon+p-p_{el}$. More importantly, this implies that the states considered in these models are not global (or even local) minima of the holographic thermodynamic free energy, \cite{Donos:2013cka, Donos:2015eew} (when the free energy is minimized, $p_{el}=0$). In spite of this, they are locally thermodynamically stable, with a positive definite static susceptibility matrix. Accordingly, they do not have poles in the upper half complex frequency plane. Their low-energy dynamics is also precisely given by the effective theory developed in the previous section, \cite{Ammon:2020xyv}.

Helical homogeneous or inhomogeneous models do not require background strain, as the free energy can be minimized non-trivially as a function of the modulation wavevector, \cite{Donos:2011ff,Donos:2012gg,Donos:2012wi,Rozali:2012es,Donos:2013wia, Withers:2013loa,Withers:2014sja,Donos:2015eew}, but are technically more challenging to work with.

The underlying conformal invariance of holographic models places a number of constraints on the equation of state and transport coefficients, since the stress-energy tensor is now traceless $T^\mu_\mu=0$. For instance, the transverse and longitudinal speeds of sound obey a simple relation, \cite{Esposito:2017qpj,Armas:2019sbe}:
\begin{equation}\label{qq}
    c_\parallel{}^2\,=\,\frac{1}{d-1}\,+2\,\frac{d-2}{d-1}\,c_\perp{}^2\,.
\end{equation}

Increasing the background strain gives additional contributions to the effective elastic moduli through \eqref{renormBG}. Depending on the specific functional dependence on strain, this may lead to thermodynamic instabilities if the effective elastic moduli vanish (this leads in turn to a divergence of the corresponding susceptibilities \eqref{GoldstoneStaticSusceptibilities}). These thermodynamic instabilities have dynamical counterparts, as e.g. the transverse sound velocity \eqref{TransverseSoundMode} becomes complex  or the longitudinal diffusive mode $D_\parallel$ \eqref{LongitudinalModes} crosses to the upper half plane. The conjectured endpoint of this instability is the nucleation of topological defects, which relax the background strain, and probably leads to a plastic behaviour and the failure of the rigidity of the system.\footnote{In this sense, this mechanism shares several commonalities with the Landau instability for superfluids triggered by a background superfluid velocity \cite{lifshitz2013statistical} and already observed in bottom-up holographic models \cite{Amado:2013aea,Lan:2020kwn}.} Viscoelasticity with background strain (or equivalently stress) has been discussed in several engineering-oriented works \cite{biot1965internal,doi:10.1063/1.1712807,doi:10.1063/1.1710417,berjamin2021acoustoelastic,destrade2007waves,destrade2009small,doi:10.1121/1.1911738}. The onset of instability in the dispersion relation of the low-energy modes has been experimentally observed in \cite{PhysRevLett.91.135501,PhysRevB.89.184111} and recently re-formulated in the context of relativistic effective field theories \cite{PhysRevD.100.065015,Pan:2021cux}. Those instabilities have not yet been investigated by holographic methods, though see \cite{Baggioli:2020qdg} for first steps with a pure shear strain.

In holographic systems, the black hole horizon provides a large bath of $\mathcal O(N^2)$ degrees of freedom. It is natural to expect that the Goldstone can relax into this bath. This is embodied by a modernized version of the ``membrane paradigm'', \cite{Damour:2008ji,1986bhmp.book.....T}, whereby transport coefficients characterizing linear response are expressed in terms of the background solution evaluated on the black hole horizon through the construction of radially conserved bulk fluxes, \cite{Iqbal:2008by},\cite{Donos:2014cya,Donos:2014yya}.

This was used to great effect to compute the linear, relativistic transport coefficients \eqref{constreldiss} in holographic models of spontaneously broken translations, either homogeneous \cite{Amoretti:2017axe,Amoretti:2019cef} or inhomogeneous, \cite{Donos:2018kkm,Gouteraux:2018wfe}. In homogeneous holographic models, it is very well understood how to encode for quantum critical infrared fixed points with broken translations, \cite{Gouteraux:2014hca, Donos:2014oha}. Near such critical phases, it was observed \cite{Amoretti:2019cef} that some of the transport coefficients are not independent and saturate a bound originating from positivity of entropy production, \eqref{entropybounds},
\begin{equation}
\label{relationstransportcoeffslowT}
    \xi_\mu=-\left(\frac{\mu}{\chi_{\pi j_q}}\right)\sigma_o\,,\quad \xi=\left(\frac{\mu}{\chi_{\pi j_q}}\right)^2\sigma_o\,,
\end{equation}
where for relativistic phases $\chi_{\pi j_q}=s_o T_o-p_{el}$. Effectively, the Goldstone relaxation processes are governed by the incoherent (i.e. without momentum drag) diffusivity $\sigma_o$, which also controls the thermal diffusivity with open circuit boundary conditions, \cite{Davison:2015taa,Davison:2018nxm}.

This can be understood as arising from the dominance of the following effective interaction between the momentum and the heat current $j_q^i$ in the infrared Hamiltonian:
\begin{equation}
\label{effectiveHlowT}
    \Delta H=\frac{1}{\chi_{\pi j_q}}\int d^dx\,\pi_i j_q^i\,.
\end{equation}
This in turn implies that 
\begin{equation}
    \dot u^i=i~[H,u^i]=\frac{j_q^i}{\chi_{\pi j_q}}\,.
\end{equation}
Plugging this in the Kubo formul\ae\ for $\xi_\mu$ and $\xi$, \eqref{Kuboximu}, \eqref{Kuboxi}, and evaluating them, leads to \eqref{relationstransportcoeffslowT}.

The reader may wonder why the specific coupling \eqref{effectiveHlowT} appears rather than some arbitrary linear combination of the electric and heat currents. It is plausible that this is an artifact of the homogeneous holographic Q-lattice/massive gravity models, where the heat current plays a distinguished role in relaxation processes, \cite{Blake:2015epa,Donos:2019hpp}. Whether this remains true in homogeneous helical models \cite{Donos:2012wi,Andrade:2017cnc} or in inhomogeneous models \cite{Rozali:2012es,Donos:2013wia, Withers:2013loa} is an open question, although the recent numerical results of \cite{Andrade:2022udb} tend to indicate a negative answer. The difference between these models is the Chern-Simons bulk term and associated breaking of parity, as well as the absence of background strain. A complete match between the hydrodynamics of the previous sections and those models is also yet to appear.\footnote{\label{footnote:PB}Probe brane models can also display spontaneous breaking of translations, \cite{Jokela:2014dba,Jokela:2016xuy}. Importantly, because of the presence of an additional long-lived mode \cite{Karch:2008,Nickel:2010pr,Davison:2011ek,Chen:2017dsy}, the hydrodynamics presented in this \textit{Colloquium} does not apply directly to these holographic models.}

%%%%%%%%%%%%%%%%%%%%%%%%%%%%%%%%%%%%%%%%%%%%%%%

\subsection{Emergent higher-form symmetries and topological defects \label{section:HF}}

Ordinary, $0$-form symmetries (such as a global U(1)) give rise to conserved, one-form currents (e.g. $\nabla_\mu J^\mu=0$ in relativistic notation). The associated conserved charges are point-like objects. 
\cite{Gaiotto:2014kfa} pointed out the existence of more general symmetries associated to differential forms of a higher rank. A prototypical example is the U(1) of electromagnetism in four spacetime dimensions. There, the Bianchi identity can be reformulated as the conservation equation of a magnetic U(1) symmetry by Hodge dualizing the Maxwell field strength $J^{\mu\nu}=1/2\,\epsilon^{\mu\nu\rho\sigma}F_{\rho\sigma}$. The charge $Q=\int_\Sigma \star J$ counts the number of magnetic lines across a codimension-$2$ surface $\Sigma$, and its associated conserved current is now a two-form $\nabla_\mu J^{\mu\nu}=0$. Among various applications, this provides a starting point for a consistent formulation of magnetohydrodynamics \cite{Grozdanov:2016tdf}.

A similar treatment can be applied to phases with a spontaneously broken global U(1) symmetry (superfluids). Keeping to relativistic notation, the absence of topological defects (vortices) implies that derivatives commute, $\nabla_{[\mu}\nabla_{\nu]}\phi=0$, where $\phi$ here is the superfluid phase. In $2+1$ dimensions, defining $J^{\mu\nu}=\epsilon^{\mu\nu\rho}\nabla_\rho\phi$ leads to an emergent conservation equation for the higher-form symmetry U(1)$_w$ associated to the conservation of winding planes, $\nabla_\mu J^{\mu\nu}=0$, \cite{Delacretaz:2019brr}. 
This emergent symmetry is broken when the theory is coupled to a background gauge field for the microscopic U(1). Indeed, the background gauge field appears as a mixed 't Hooft anomaly on the right-hand side of the conservation of the U(1)$_w$, $\nabla_\mu J^{\mu\nu}=-q \epsilon^{\nu\kappa\lambda}F_{\kappa\lambda}$ where $F_{\kappa\lambda}=\nabla_{[\kappa}A_{\lambda]}$ and the anomaly coefficient $q$ is the charge of the condensate.

Emergent symmetries are often anomalous and their higher-form generalizations are no exception, \cite{Gaiotto:2014kfa,Landry:2021kko}. Such anomalies give rise to anomaly matching conditions, which put strong constraints on the hydrodynamic gradient expansion. Ultimately, for superfluids they are responsible for the emergence of second sound, whose velocity is proportional to the anomaly coefficient, \cite{Delacretaz:2019brr}, and give rise to dissipationless transport, \cite{Else:2021}. Preliminary investigations of the higher-form symmetry formulation\footnote{This differs from the dual formulations of \cite{Beekman:2016szb,Beekman:2017brx}.} of phases with spontaneously broken translations can be found in \cite{Grozdanov:2018fic,Armas:2019sbe}, but do not include any mixed `t Hooft anomaly. Whether the anomaly-based mechanism for sound modes and dissipationless transport also operates for spacetime symmetries remains to be understood.

In the condensed phase, the winding operators $W_{ij}=\int d^2 x\nabla_i u_j$ are conserved and measure elastic deformations of the crystal/density wave. They lead to undamped propagation of uniform bulk and shear strains, e.g.\footnote{Notice the difference with the Kubo formula in Eq.\eqref{thiseq} in which the divergent $1/\omega$ term would not appear.}
\begin{equation}
\label{ShearCond}
    \eta(\omega)\equiv\frac{i}{\omega} G^R_{\tau^{xy}\tau^{xy}}(\omega,q=0)=\eta+G~\frac{i}{\omega}\,.
\end{equation}
This infinite dc `shear conductivity' is the analogue of dissipationless charge transport in superfluids.

At finite temperatures, bound pairs of defects/anti-defects (dislocations or disclinations) nucleate. Above the Berezinsky-Kosterlitz-Thouless temperature, thermal fluctuations lead to their unbinding and they become mobile -- the BKT phase transition, \cite{Kosterlitz_1973,PhysRevLett.41.121,PhysRevB.19.2457,PhysRevB.22.2514}.\footnote{See \cite{Kivelson1998,PhysRevLett.108.267001,PhysRevB.86.115138,Beekman:2016szb,Beekman:2017brx} for the quantum case.} Mobile defects relax the windings, and the corresponding emergent symmetry U(1)$_w$ is explicitly broken.\footnote{The explicit breaking of higher form symmetries has been considered using effective field theory methods in \cite{Baggioli:2021ntj}.} This leads to relaxation of the longitudinal and transverse phonons, \begin{equation}
    \dot\lambda_{\parallel,\perp}=-\Omega_{\parallel,\perp}\lambda_{\parallel,\perp}+\dots
\end{equation} 
This equation is valid when the anisotropic rates $\Omega_\parallel$ and $\Omega_\perp$ are small, close to the BKT phase transition. The phase relaxation rates are set by the viscosities of the normal phase and the density of free defects $n_f$, e.g. $\Omega_\perp\sim n_f/\eta_{normal}$, \cite{PhysRevB.22.2514}.\footnote{See \cite{Delacretaz:2017zxd} for a memory matrix calculation of these rates.} `Climb' motion of dislocations is usually suppressed compared to `glide', i.e. $\Omega_\parallel\ll\Omega_\perp$. In the language of higher-form symmetries, the emergent higher-form symmetry counting winding planes is broken by irrelevant operators (the defects), \cite{Delacretaz:2019brr}.

Evaluating \eqref{ShearCond} again, the viscosities of the condensed phase are finite but large, $\eta(\omega=0)=\eta+G/\Omega_\perp$.

%%%%%%%%%%%%%%%%%%%%%%%%%%%%%%%%%%%%%%%%%%%%%%%
\section{Pseudo-spontaneous breaking of translations}
\label{sec:3}

The total momentum of the full system is always conserved, due to the translation invariance of the ambient spacetime in which the crystal lives. Thus, the emergent continuous translation symmetry at long distances in crystalline solids cannot be explicitly broken. In systems at finite density such as metals, the conduction electrons (or more generally the charge carriers at strong coupling) can be considered in some regimes (typically, low enough temperatures) to be weakly-coupled to lattice degrees of freedom and other sources of inelastic scattering. The electron momentum then becomes approximately conserved, with an emergent electronic translation symmetry in the infrared broken by irrelevant operators (such as umklapp). Disorder gives rise to elastic scattering and to a residual zero temperature resistivity, and so should be weak in order for momentum to remain approximately conserved. In an electronic charge density wave or Wigner crystal phase, electronic translations are spontaneously broken and give rise to a spatially modulated electronic density of states (see \cite{RevModPhys.60.1129} for a review). New Goldstone degrees of freedom emerge, called {phasons} or sometimes phonons by abuse of terminology (not to be confused with the phonons of the underlying lattice).

It then becomes interesting to study how the weak explicit breaking mentioned above affects the dynamics of the Goldstones. These acquire both a small mass $q_o$\footnote{By a similar mechanism that leads to the Gell-Mann Oakes Renner (GMOR) relation \cite{PhysRev.175.2195} for pion masses in QCD.} and a damping $\Omega$, leading to a nonzero real and imaginary part in their $q=0$ dispersion relation, respectively. Phenomenologically, the spontaneous, spatially modulated phase is no longer free to slide and is pinned at a frequency $\omega_o\sim c_s q_o$ proportional to the mass of the Goldstone -- which now has a finite correlation length. Correspondingly, there is a gap in the real part of the the frequency-dependent conductivity with a peak at a frequency $\omega\sim\omega_o$, representing the energy cost to de-pin the density wave.\footnote{If disorder or lattice effects are strong, the density wave is strongly-pinned and locked at impurity sites.}

Pinning of charge density waves is an old subject, \cite{LEE19931001, PhysRevB.17.535}, and was confirmed in many experiments on quasi one-dimensional materials, \cite{RevModPhys.60.1129}. It was revived in recent years, spurred on by a combination of mounting experimental evidence on the role of charge density wave phases or fluctuations across the phase diagram of cuprate high $T_c$ superconductors,  \cite{Peng:2018,Arpaia:2019,Miao:2021,Lin:2021,Lee:2020,Ma:2021,Tam:2021,Kawasaki:2021,Lee:2021generic} (see \cite{Arpaia:2021} for a review); theoretical developments on the application of hydrodynamics and related effective field theoretic descriptions of transport to strongly-correlated electronic materials, \cite{Hartnoll:2014lpa,Lucas:2015pxa,Levitov:2016,Zaanen:2018edk}; and the development of holographic methods for phases with broken translations.

 Following the initial work of \cite{Delacretaz:2017zxd}, which incorporated pinning by explicit breaking of translations and damping by defects into a hydrodynamic framework, a number of groups set out to investigate these phases using holographic methods. The original expectation was that these systems would display a pinning frequency $\omega_o$ and a momentum relaxation rate $\Gamma$, but no phase relaxation rate $\Omega$, as none of these holographic models included mobile elastic defects.\footnote{Though see \cite{Andrade:2017ghg,Krikun:2017cyw} for a holographic construction of phases with static discommensurations.} It then initially came  as a surprise when it was recognized that they exhibited a finite phase relaxation rate governed by the pseudo-Goldstone mass and diffusivity $\Omega =Gq_o^2 \xi$, \cite{Amoretti:2018tzw,Donos:2019txg} with further confirmations in \cite{Andrade:2018gqk,Ammon:2019wci,Donos:2019hpp,Baggioli:2019abx,Amoretti:2019kuf,Andrade:2020hpu}. 
 
 It is worth noting that the assumption of hydrodynamics is not necessary to the existence of a phase relaxation rate $\Omega\sim\omega_o^2$ in the presence of weakly-broken translations. A memory matrix approach (see \cite{forster,Hartnoll:2016apf} for reviews) suffices, \cite{Delacretaz:2017zxd}. Where hydrodynamics enters is in the determination of the relevant memory matrix element in terms of a diffusive transport coefficient $\xi$. This belongs to the same class of hydrodynamic relaxation mechanisms giving rise to flux-flow resistance in phase-relaxed superconductors \cite{PhysRev.140.A1197,Davison:2016hno} or minimal viscosity scenarii for cuprate strange metals, \cite{Davison:2013txa,Zaanen:2018edk}. As we elaborate upon below, $\Omega$ captures the contribution of ungapped excitations to the dc resistivity.

The main theoretical achievement of this collective effort is the construction of a hydrodynamic theory of pseudo-spontaneously broken translations, \cite{Delacretaz:2021qqu,Armas:2021vku}, which explains the observations above and which we now describe. For simplicity we will consider states without background strain throughout this section, but this can be incorporated straightforwardly,  \cite{Armas:2021vku}.

\subsection{Hydrodynamics}

 When translations are weakly broken explicitly, the free energy at quadratic order in fluctuations now includes a mass term for the Goldstone modes\footnote{The mass term can be thought to originate from expanding a $\cos u_i$ deformation of the Hamiltonian of the system to quadratic order in fluctuations, so the $u_i$ are still compact scalars.}
\begin{equation}\label{freeenergypinning}
    \delta f^{(2)}=\frac{B+G}2\left(\nabla^i\delta\phi_i\right)^2+\frac{G}2\left(\nabla\times\delta\phi\right)^2+\frac{G \,q_o^2}{2}\,\delta\phi_i\delta\phi^i
\end{equation}
which shifts the unpinned static susceptibility matrices $\chi_{o,\parallel}$ and $\chi_{o,\perp}$ as
\begin{equation}
    \chi_o^{-1}\mapsto\chi^{-1}=\chi_{o}^{-1}+\Delta\chi^{-1}\,,
\end{equation}
where $\Delta\chi^{-1}$ is a matrix whose only nonzero elements are $(\Delta\chi^{-1})_{\lambda_{\parallel}\lambda_{\parallel}}=(\Delta\chi^{-1})_{\lambda_{\perp}\lambda_{\perp}}=G\, q_o^2/q^2$. As a result, the static susceptibility matrix $\chi$ becomes nonlocal. 

The charge and energy conservation equations in \eqref{hydroeoms} remain unchanged. On the other hand, since translations are broken explicitly, momentum is no longer conserved
\begin{equation}
    \dot\pi^i+\nabla_j\tau^{ji}=-\Gamma \pi^i-Gq_o^2\delta\phi^i\,.
\end{equation}
The $\Gamma$ term is allowed on general grounds and captures momentum relaxation, while the second term encodes the effects of the mass of the Goldstone, and can be derived by computing $\dot\pi^i=i[H,\pi^i]$ including a mass deformation \eqref{freeenergypinning} in the Hamiltonian $H$ and using the commutator \eqref{commutatorgoldstone}.

The constitutive relations and the Josephson equation can all contain terms linear in $\phi^i$ without any spatial gradient, since the shift symmetry is broken. These terms are constrained by locality and Onsager relations. After imposing these constraints, the constitutive relations and the Josephson equation read
\begin{equation}\label{constrelpinned}
    \begin{split}
        j^i=&-\sigma_o\nabla^i\mu-\alpha_o\nabla^i T+\xi_\mu h^i\,,\\
        \frac{\tilde j_q^i}T&=-\alpha_o\nabla^i\mu-\frac{\bar\kappa_o}{T}\nabla^i T+\xi_T h^i\,,\\
        \tau^{ij}=&-\eta\,\sigma^{ij}-\zeta\nabla\cdot v\delta^{ij}\,,\\
        \dot\phi^i=& v^i+\xi_\mu\nabla^i\mu+\xi_T\nabla^i T-\xi  \,h^i
    \end{split}
\end{equation}
where $h^i=\partial f/\partial \phi^i=G q_o^2 \delta\phi^i-\nabla_j h^{ji}$ and in the absence of background strain the transport coefficients are no longer matrices. These dissipative corrections ensure that the equations of motion remain local \cite{Delacretaz:2021qqu} and that the divergence of the entropy current is positive, \cite{Armas:2021vku}. Translating the $h^i$ terms to fields $\phi^i$ generates new relaxation terms in the constitutive relations and Josephson equations, proportional to $q_o^2$ and various dissipative transport coefficients; $\xi_\mu$, $\xi_T$ and $\xi$. For instance, the Josephson equations take the form
\begin{equation}
    \dot\delta\phi^i=-\Omega\,\delta\phi^i+O(\nabla^i)
\end{equation}
where the damping term $\Omega$
\begin{equation}
    \Omega=Gq_o^2\xi \label{uni}
\end{equation}
is universally determined by the Goldstone mass and $\xi$. The parameter $\xi$ is a diffusive transport coefficient of the translation invariant theory which enters in the attenuation of sound and diffusive modes of section \ref{sec11} and encodes dissipation of the Goldstone mode in the thermal bath over long distances.

 In the framework of effective (hydrodynamic) theories, \eqref{uni} is a direct consequence of locality \cite{Delacretaz:2021qqu} or the second law of thermodynamics \cite{Armas:2021vku} with external sources on.\footnote{Analogous relations apply for other symmetry-broken phases, such as superfluids \cite{Ammon:2021slb,Delacretaz:2021qqu,Armas:2021vku}, QCD in the chiral limit \cite{Grossi:2020ezz,Grossi:2021gqi}, nematic phases, (anti-)ferromagnets \cite{Delacretaz:2021qqu} and quasicrystals \cite{Baggioli:2020nay,Baggioli:2020haa}.} 

The damping term $\Omega$ is allowed on general grounds, since the shift symmetry of the Goldstones is broken by the explicit breaking of translations, without having to assume a hydrodynamic regime. A  memory matrix analysis, \cite{Delacretaz:2017zxd}, shows that it is given by the following Kubo formula
\begin{equation}
    \Omega=Gq_o^2\lim_{\omega\to0}\frac1\omega\textrm{Im}G^R_{\partial_t\phi^i\partial_t\phi^i}(\omega,k=0)\,,
\end{equation}
In this approach, the retarded Green's function on the right-hand side should be evaluated in the purely spontaneous theory. Using the hydrodynamic form of the retarded Green's function gives back \eqref{uni}.

In the presence of explicit breaking, $\Omega$ captures the relaxation of the pseudo-Goldstone mode in the surrounding bath of thermal excitations. In \cite{Andrade:2020hpu}, it was shown that in the absence of a gap the time-dependent Ginzburg-Landau equation gives a good account of the dynamics of these systems near $T_c$. For one-dimensional systems with quasi-perfect nesting of the modulation wavevector and gapping of the Fermi surface, the charge density wave formation is described by the Peierls instability, \cite{RevModPhys.60.1129}. The gap equation is BCS-like and the density of uncondensed electrons is exponentially suppressed at low temperatures. In this case, there are very few thermal excitations that the pseudo-Goldstone can relax into and we expect the damping $\Omega$ to be suppressed, which explains why it has not been discussed in previous literature, \cite{RevModPhys.60.1129}. In other words, in the absence of a thermal bath, the Goldstone mode is gapped and cannot `leak' to arbitrarily low energies.

Pinning also introduces new relaxation parameters in the constitutive relations for the currents
\begin{equation}
    j^i=nv^i+\Omega_n\delta\phi^i+O(\nabla)\,,\quad \frac{j_q^i}T=s v^i+\Omega_s \delta\phi^i+O(\nabla)
\end{equation}
with 
\begin{equation}
\label{Omegans}
    \Omega_n=G q_o^2 \xi_\mu\,,\quad \Omega_s=G q_o^2\xi_T\,.
\end{equation}

With translations broken explicitly weakly, the quasi-normal modes of the system have both an imaginary and a real gap
\begin{equation}\label{pinneddisprel}
    \omega_\pm=\pm \sqrt{\frac{G}{\chi_{\pi\pi}}}q_o-\frac{i}2\left(\Gamma+G q_o^2\xi\right)+O\left(q^2,g^3\right)\,.
\end{equation}
In the equation above, we have assumed the scaling $q_o\sim g$, $\Gamma\sim g^2$, where $g$ is the source of the microscopic operator breaking translations explicitly. This assumption can be lifted, and then the dispersion relation takes a more complicated form. The expression \eqref{pinneddisprel} makes manifest the damped oscillator behavior of the system, with a pinning frequency $\omega_o\equiv q_o\sqrt{G/\chi_{\pi\pi}}$, and two contributions to the damping rate: $Gq_o^2\xi$ takes a universal form in terms of parameters of the effective field theory, while $\Gamma$ does not.
The only gapless modes left are two diffusive modes transporting charge and thermal fluctuations. Their expressions, as well as the leading $q$-dependence of the gapped modes can easily be computed with \eqref{constrelpinned} in hand, but their expressions are not particularly illuminating and we leave it to the interested reader to write them down.

In \cite{Armas:2021vku}, extra coefficients have been reported when coupling to external sources. Since these terms originate from extra freedom in how currents are coupled to external sources when symmetries are explicitly broken, they only appear in the numerator of retarded Green's functions and do not affect the poles. In particular, they do not affect the relations \eqref{uni} and \eqref{Omegans}. It is also not known at the time of writing this review how they affect the electric conductivity, which will be our primary focus in the next section. For simplicity, we will then omit these terms and refer to \cite{Armas:2021vku} for details. This is justified to some extent by the fact that these terms are either absent from or can be redefined away in the holographic models with pseudo-spontaneous breaking investigated so far, \cite{Amoretti:2018tzw,Donos:2019hpp,Donos:2019tmo,Donos:2019txg,Ammon:2019wci,Ammon:2021slb}. 

%%%%%%%%%%%%%%%%%%%%%%%%%%%%%%%%%%%%%%%%%%%%%

\subsection{Charge transport in pinned crystals\label{section:chargetransport}}

In a translation-invariant system at nonzero density, the electric conductivity is infinite in the dc limit $\sigma_{dc}\equiv\sigma(\omega=0)$. This is because at nonzero density the electric current, which is a fast mode, overlaps with the (electronic) momentum density, which is conserved. This is manifested in a nonzero cross-susceptibility $\chi_{JP}$ between the charge and momentum operators. Hence the electric current cannot relax, which manifests itself as a divergence of the zero frequency conductivity. This can be proven rigorously on general grounds using the memory matrix formalism, see e.g. \cite{Hartnoll:2016apf}. This continues to be true when translations are spontaneously broken (e.g. for an electronic charge density wave in a clean system). Using the hydrodynamics equations of the previous sections, the conductivity can be obtained from the Ward identity for charge conservation:\footnote{While in hydrodynamics the continuity equation is a dynamical equation for the time evolution of vevs of operators, the Ward identity is a consequence of the U(1) symmetry and is more fundamental. It is an operator equation which can be used inside Green's functions.}
\begin{equation}
    \sigma(\omega)\equiv\frac{i}\omega G^R_{jj}(\omega,q=0)=\frac{i}\omega\lim_{q\to0}\frac{\omega^2}{q^2}G^R_{nn}(\omega,q)
\end{equation}
and is found to be
\begin{equation}
    \sigma(\omega)=\sigma_o+\frac{n_o^2}{\chi_{\pi\pi}}\frac{i}\omega\,.
\end{equation}
The $\omega=0$ pole in the imaginary part is physical and cannot be removed by contact terms. As announced, its residue is directly proportional to the off-diagonal susceptibility $\chi_{JP}=n_o$, which is identified as the charge density of the system. It gives rise to a delta function in the real part through Kramers-Kr\"onig relations. There is also a finite contribution to the real part, captured by the transport coefficient $\sigma_o$. It is always nonzero except in Galilean-invariant system, where it vanishes as a consequence of the Ward identity for Galilean boosts, $j^i=\pi^i$ (where for simplicity we set the electric charge and particle mass to unity in this formula). Intuitively, it is the contribution to electric transport of `incoherent' processes (meaning which do not give rise to dissipationless current) \cite{Davison:2015taa}. It has no equivalent in a simple quasi-particle picture, which is intrisically Galilean-invariant. It can be generated in Boltzmann transport by including terms breaking Galilean invariance, see for instance, \cite{Huang:2020uvj}. It would also be present in a translation-invariant fluid without Galilean boosts, and there it transports fluctuations of entropy per unit charge $\delta(n/s)$ diffusively, \cite{Hartnoll:2016apf} (when translations are spontaneously broken, the eigenmode is more complicated due to the coupling to the longitudinal Goldstone).

When translations are explicitly broken, the electronic momentum is no longer conserved. In the regime where it relaxes slowly enough to be kept in the effective field theory as a light mode, the conductivity is strongly modified. It is helpful to first consider the case without spontaneous breaking, \cite{Hartnoll:2007ih}. The only relaxation parameter is the momentum relaxation rate $\Gamma$, and the electric conductivity becomes\footnote{This formula is really only correct to order $\mathcal O(1/\Gamma)$. Generally, susceptibilities will receive $\mathcal O(\Gamma)$ corrections which need to be included in order to consistently describe the dc conductivity to order $\mathcal O(\Gamma^0)$, \cite{Davison:2015bea}.}
\begin{equation}
\label{sigmaGamma}
    \sigma(\omega)=\sigma_o+\frac{n_o^2}{\chi_{\pi\pi}}\,\frac1{\Gamma-i\omega}+\mathcal O(\Gamma^0)\,.
\end{equation}
The $\omega=0$ pole is now located at $\omega=-i\Gamma$ and is identified with slowly-relaxing momentum. In real space, we expect $\langle\pi^i(t)\rangle\sim\pi^i_0 e^{-\Gamma t}$. The real part of the conductivity shows a sharp peak centered at zero frequency (the Drude peak), of width $\Gamma$ and weight $n_o^2/(\chi_{\pi\pi}\Gamma)$. In the weakly-relaxing regime, $\Gamma\ll\Lambda$ ($\Lambda$ being the thermalization scale), the dc conductivity $\sigma_{dc}=\sigma_o+n_o^2/(\chi_{\pi\pi}\Gamma)\simeq n_o^2/(\chi_{\pi\pi}\Gamma)$ is large and completely dominated by this `Drude' contribution. The system is a hydrodynamic metal where the electronic momentum relaxes by inelastic scattering off impurities or by Umklapp processes. 

\begin{figure}
\includegraphics[width=.47\textwidth]{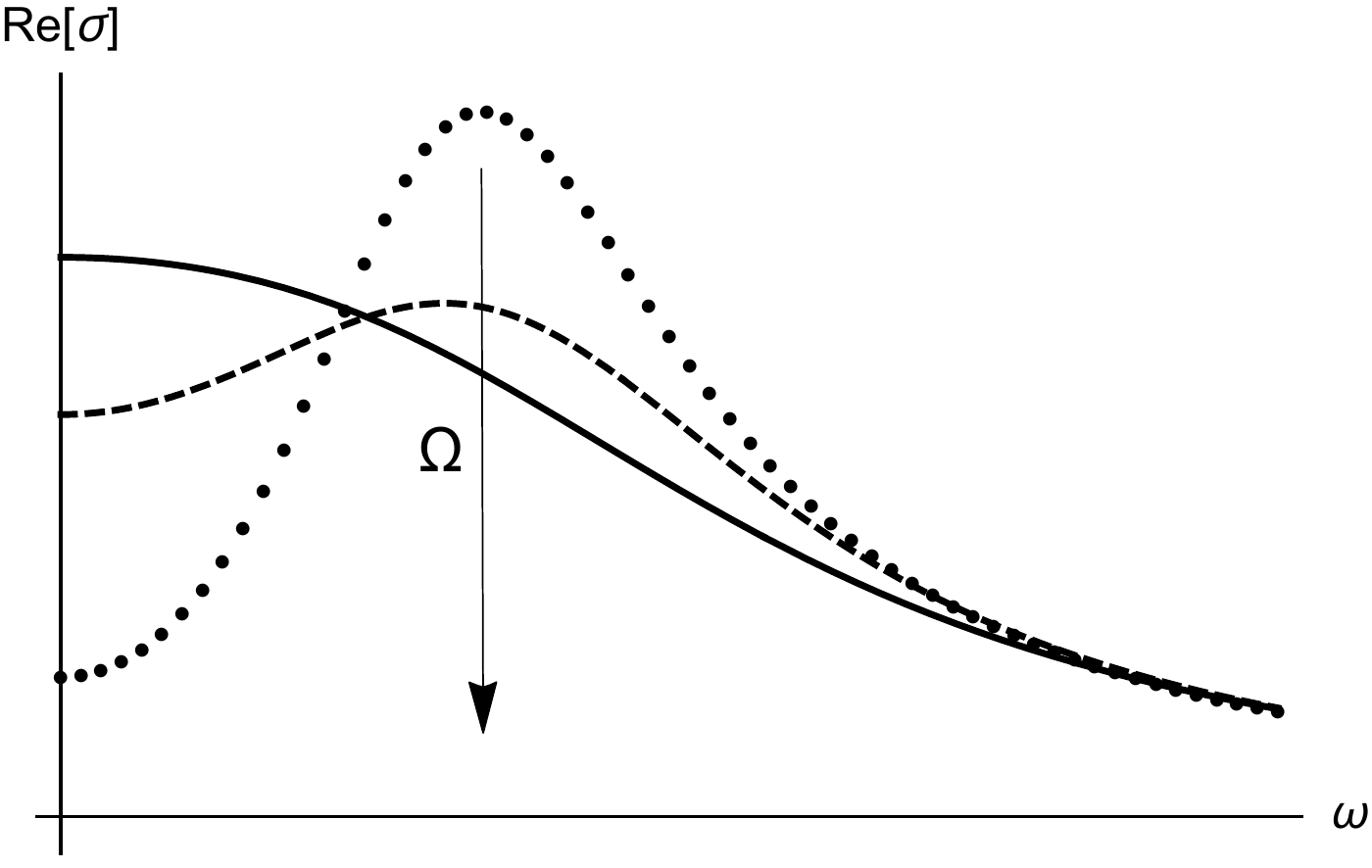}
\caption{\label{fig:ReSigmaOmega}Schematic representation of the ac conductivity \eqref{accondpinned} in the Galilean limit when $\Omega$ is dialed from large (solid line) to small (bullets), keeping all other parameters fixed. The transfer of spectral weight from the zero frequency Drude-like peak to the off-axis peak is evident.}
\end{figure}

By contrast, when translations are pseudo-spontaneously broken, the frequency-dependent conductivity becomes
\begin{equation}
\label{accondpinned}
    \sigma(\omega)=\sigma_o+\frac{\frac{n_o^2}{\chi_{\pi\pi}}(i\omega-\Omega)+2n_o\Omega_n+\frac{\Omega_n^2}{\omega_o^2}(\Gamma-i\omega)}{(\omega+i\Gamma)(\omega+i\Omega)-\omega_o^2}\,.
\end{equation}
Compared to the case without spontaneous breaking of translations, we observe new contributions to inelastic scattering, proportional to $q_o^2$ and contained in the $\omega_o$, $\Omega$ and $\Omega_n$ terms. The lineshape interpolates between a Lorentzian centered at $\omega_o$ when $\Omega$ and $\Omega_n$ can be neglected (matching previous hydrodynamic treatments of the collective zero mode, \cite{RevModPhys.60.1129}), and a Drude-like peak centered at $\omega=0$ when the damping rates become more important, as is illustrated in figure \ref{fig:ReSigmaOmega} (for the original argument see \cite{Delacretaz:2016ivq}). In \cite{RevModPhys.60.1129} and experimental references therein, the focus was on low temperatures. It would be interesting to confront the formula \eqref{accondpinned} to experimental data at higher temperatures, where ungapped degrees of freedom become non-negligible.

The dc conductivity 
\begin{equation}
\label{dcconductivity}
    \sigma_{dc}=\sigma_o+\frac{\frac{n_0^2}{\chi_{\pi\pi}}\Omega-2n_o\Omega_n-\frac{\Omega_n^2}{\omega_o^2}\Gamma}{\Gamma\Omega+\omega_o^2}
\end{equation}
is non-vanishing due to the nonzero symmetry-breaking terms $\Omega$, $\Omega_n$ and to the `non-Galilean' transport coefficient $\sigma_o$. Previous hydrodynamic treatments (see e.g. \cite{RevModPhys.60.1129}) usually assume the Galilean limit, where the coefficients $\sigma_o$, $\xi_\mu$ and consequently $\Omega_n$ would be zero, but did not account for $\Omega$ in the dynamics of the collective mode. Here, in the Galilean limit (setting $n_o=n e$ and $\chi_{\pi\pi}=m n$, where $e$ is the electron unit charge, $n$ the density and $m$ the mass), the resistivity $\rho\equiv1/\sigma(\omega=0)$ is
\begin{equation}
\label{resgal}
    \rho_{Galilean}=\frac{m}{n e^2}\Gamma+\frac{1}{(ne)^2\xi}\,.
\end{equation}
In other words, we do not expect a translation-broken phase such as a charge density wave to be necessarily insulating: the inelastic scattering of the Goldstone into the bath of thermal excitations provides a conduction channel. In electronic charge density wave materials, such as those reviewed in \cite{RevModPhys.60.1129}, the Fermi surface may only be partially gapped in the charge density wave phase. The second term in \eqref{resgal} captures the inelastic scattering of the Goldstone mode into these uncondensed electrons. Instead, if the Fermi surface is fully gapped, $\Omega=0$ and the collective mode does not contribute to dc transport, as in \cite{RevModPhys.60.1129}. Said otherwise, in the absence of gapless thermal excitations, there is a finite energy cost to make the density wave slide. 

The resistivity \eqref{resgal} takes a Drude-like form with a `transport scattering rate' 
\begin{equation}
    \frac1{\tau_{tr}}=\Gamma+\frac{1}{(mn)\xi}\,.
\end{equation}
In general, this picture is misleading, as there is no single pole located at $\omega=-i/\tau_{tr}$ as in the Drude model \eqref{sigmaGamma}. Rather, both poles \eqref{pinneddisprel} give important contributions to the lineshape. When the Goldstone damping rate is large compared to $\Gamma$ and $\omega_o$, the poles are located at
\begin{equation}
    \omega_-=-i\left(\Omega-\frac{\omega_o^2}{\Omega}+\dots\right)\,,\quad \omega_+=-i\left(\Gamma+\frac{\omega_o^2}{\Omega}+\dots\right),
\end{equation}
where the dots denote subleading $1/\Omega$ corrections. The $\omega_-$ pole recedes deep in the lower-half plane and drops out of the effective theory, while the $\omega_+$ one remains long-lived. Accordingly, the ac conductivity becomes Drude-like, as in the blue solid line of figure \ref{fig:ReSigmaOmega}. In \cite{Amoretti:2018tzw}, this process takes place at very low temperatures.

Above $T_c$, we also expect to recover a Drude-like conductivity as in \eqref{sigmaGamma}, dominated by a single pole $\omega\simeq -i\Gamma$. In \cite{Andrade:2020hpu}, it was shown that this occurs through a vanishing of the residue of the $\omega_-$ pole, while $\Omega$ remains finite through the phase transition. Above $T_c$, there is no condensate but $\Omega$ captures the fluctuations of the order parameter. 

Moving away once again from the Galilean limit, the dc conductivity \eqref{dcconductivity} no longer depends on the explicit symmetry-breaking parameter $q_o$ after inserting the relations \eqref{uni} and \eqref{Omegans}. At low temperatures, inelastic scattering off impurities is expected to dominate the momentum relaxation rate, and does not contribute any temperature dependence $\Gamma\sim T^0$. Then, the primary temperature dependence of the resistivity originates from incoherent scattering processes encapsulated in the diffusive transport coefficients $\sigma_o$, $\xi_\mu$ and $\xi$, in sharp contrast to metallic phases. In a metal, extrinsic processes dominate the resistivity through the scattering rate, $\rho_{dc}\sim\Gamma$, while when translations are pseudo-spontaneously broken, intrinsic ones do.

%%%%%%%%%%%%%%%%%%%%%%%%%%%%%%%%%%%%%%%%%%%%%%%
\subsection{Holography\label{section:HoloSP}}

Pseudo-spontaneous breaking of translation symmetry has been implemented in several holographic models in the past years \cite{Ling:2014saa, Amoretti:2016bxs,Andrade:2017cnc,Jokela:2017ltu,Alberte:2017cch,Andrade:2017ghg,Li:2018vrz,Andrade:2018gqk,Amoretti:2018tzw,Donos:2019tmo,Ammon:2019wci,Baggioli:2019abx,Donos:2019hpp,Amoretti:2019kuf,Andrade:2020hpu}. Independently of the concrete model at hand, this limit is always achieved by introducing on top of the purely spontaneous state a small space-dependent source for a boundary operator which is therefore responsible for the explicit translation symmetry breaking.\footnote{In some of the examples, this boundary operator is the same which breaks translations spontaneously.}

This body of work firmly established the validity of the hydrodynamic theory of pseudo-phonons presented above \footnote{For the reasons mentioned in footnote \ref{footnote:PB}, the hydrodynamics of probe brane setups are of a different nature.} and more specifically of the relation \eqref{uni}. Recent works \cite{Delacretaz:2021qqu,Armas:2021vku} further confirmed that this relation is not an artifact either of the homogeneity of the holographic models used or of the large $N$ limit inherent to the holographic approach. Various works also established a GMOR-like relation between the mass of the pseudo-Goldstone, the condensate and the source of explicit breaking, \cite{Andrade:2017cnc,Li:2018vrz,Andrade:2018gqk,Amoretti:2018tzw,Ammon:2019wci,Baggioli:2019abx,Andrade:2020hpu,Wang:2021jfu}.\footnote{The GMOR relation itself was shown to hold in holographic QCD models, see e.g. \cite{Erlich:2005qh}, and in holographic models where a $U(1)$ global symmetry is pseudo-spontaneously broken, \cite{Argurio:2015wgr}.}

Holographic models can easily account for phases which are either insulating, where the resistivity diverges towards low temperatures, \cite{Andrade:2017cnc,Andrade:2017ghg}, or metallic, \cite{Amoretti:2018tzw}, with a vanishing resistivity at low temperatures. The former case is in some respects more similar to conventional charge density wave systems, in the sense that a gap forms and the damping rate $\Omega$ does not make a large contribution to the dc conductivity, as evidenced from the negligible value of the dc conductivity compared to the height of the off-axis peak in the ac conductivity, \cite{Andrade:2018gqk}. An important difference is that the gap is algebraic, and the resistivity diverges like a power-law.  In the helical, homogeneous setup of \cite{Andrade:2018gqk}, this scaling is rooted in the critical behavior of the infra-red geometry, in the near-horizon, near-extremal limit. Indeed, as is well-known in holographic models, such critical geometries leave a strong imprint on the scaling of transport observables at low temperatures, \cite{Gouteraux:2014hca,Donos:2014oha}. It is then surprising that the resistivity continues to scale in the inhomogeneous construction of \cite{Andrade:2017ghg}, even though there is no evidence so far of scaling behavior in the geometry. A better understanding of this result remains an open question. 

In the metallic case, an inverse transfer of spectral weight is observed \cite{Amoretti:2018tzw} as the off-axis peak in the ac conductivity smoothly interpolates back to a Drude-like peak at zero frequency upon lowering the temperature, as depicted in figure \ref{fig:ReSigmaOmega}. This is accompanied by a non-trivial motion of the poles in the lower half complex frequency plane. At low enough temperature, the poles are once again purely imaginary, and the width of the Drude-like peak is controlled by the pole closest to the real axis. Its partner quickly recedes down the axis and becomes incoherent. Whether this behavior can be reproduced in a more realistic, inhomogeneous state is not known. Nonetheless, it bears intriguing resemblance to what is observed experimentally in cuprate high $T_c$ superconductors and many other strongly-correlated materials, as we will describe in section \ref{outlook:cuprates}. 

Given that $\Omega=Gq_o^2\xi$ and $\Omega_n=G q_o^2\xi_\mu$, the same effective interaction we described in section \ref{section:HoloSp} around \eqref{effectiveHlowT} operates near homogeneous holographic quantum critical phases with pseudo-spontaneously broken translations. This further implies that the low-temperature resistivity is controlled by a single, diffusive transport coefficient $\sigma_o$ of the clean state, with subleading contributions from explicit symmetry breaking (assuming disorder and umklapp processes to be irrelevant and/or contribute no significant temperature dependence to the momentum relaxation rate), $\rho_{dc}\simeq(sT/\mu n_o)^2/\sigma_o+\mu\Gamma/n_o$. As the transport coefficient $\sigma_o$ can be computed in terms of data at the black hole horizon, it is sensitive to the scaling properties of the low temperature critical phase, and hence so is the resistivity. This does not suffice to explain the results of \cite{Andrade:2017ghg} but resonates with the scaling form of the low temperature resistivity uncovered there.

 Recently, \cite{Andrade:2022udb} investigated numerically the thermoelectric ac conductivities in helical and inhomogeneous models, and found that the numerical data can be fitted to the hydrodynamic formul\ae\ very well. Their fit allows to determine the tranport coefficients $\sigma_o$, $\xi_\mu$ and $\xi$, which they do not find obey the relations \eqref{relationstransportcoeffslowT}. The models they use break parity due to the presence of Chern-Simons terms in the bulk, and their ground states have different critical properties than homogeneous Q-lattices as well as no background strain. Which is the essential feature giving rise to \eqref{relationstransportcoeffslowT} remains an open question, especially since similar relations appear to hold in experimental realizations of Wigner solids, see the following section \ref{sec:4}.
 
In underdoped cuprates, the Hall (the dc electric transverse response in a magnetic field) and the Seebeck coefficients (the dc electric response to a temperature gradient) change sign at low temperatures, \cite{Badoux2016,PhysRevB.103.155102}, which is usually attributed to the reconstruction from a large, hole-like Fermi surface to small, electron-like pockets \cite{Doiron-Leyraud2007,PhysRevLett.100.047003,Vignolle2008} due to the formation of a CDW. An important outcome of the analysis in \cite{Andrade:2022udb} is that the Seebeck coefficient changes sign at low temperatures, without the presence of a Fermi surface or reconstruction thereof. 

When the spontaneous spatially modulated structure is coupled to an explicit lattice, one expects their periodicities to become commensurate for sufficiently large lattice strength. This phenomenon is beyond homogeneous constructions, \cite{Andrade:2015iyf}. Instead, more realistic inhomogeneous constructions display commensurability effects \cite{Andrade:2017leb}. The black hole horizon is strongly spatially modulated by the spontaneous structure, which is weak in the ultraviolet near the boundary (since it is not sourced) but important in the infrared. The explicit lattice is strong in the ultraviolet but irrelevant (weak) in the infrared. The commensurability that develops between these two structures is a reflection of a strong UV-IR mixing upon increasing the UV lattice strength and turns the system into a Mott insulator, \cite{Andrade:2017ghg}, albeit with an algebraic rather than exponential gap and reminiscent of underdoped cuprates.
%%%%%%%%%%%%%%%%%%%%%%%%%%%%%%%%%%

\section{Magnetic fields}
\label{sec:4}
External magnetic fields are a valuable experimental probe in the study of strongly correlated electronic phases of matter. They are particularly important in the context of two-dimensional systems in which they produce a rich collection of new physical phenomena \cite{Chen2005QuantumSO}. Moreover, the interplay between translational symmetry breaking and the presence of an external magnetic field results in a complex structure of low-energy excitations, including the appearance of a type-II Goldstone boson with quadratic dispersion $\mathrm{Re}(\omega)\sim k^2$ -- the {magnetophonon}. This mode arises from the hybridization of the original longitudinal and transverse phonons into a gapless magnetophonon and a gapped magnetoplasmon, which is now allowed because of the time reversal symmetry breaking induced by the magnetic field \cite{Watanabe:2014fva}. As a consequence, the original Goldstone modes are not anymore independent, $ \left[\phi_i,\phi_j\right]\neq 0$.  Following the Watanabe-Brauner argument \cite{Watanabe:2011ec,PhysRevLett.108.251602,PhysRevLett.110.091601}, the number of Goldstone modes is reduced and their dispersion converted into a quadratic type. 

Early accounts of the dynamics of two-dimensional pinned charge density waves in presence of an external magnetic field are given in \cite{PhysRevB.18.6245,PhysRevB.46.3920}. Their hydrodynamics were revisited recently in \cite{Delacretaz:2019wzh,Amoretti:2021fch,Delacretaz:2021qqu}. In the presence of pinning and a magnetic field, new relations of the type \eqref{uni} arise.

\cite{Delacretaz:2019wzh} considered the match between the hydrodynamic ac conductivity and experimental measurement in GaAs heterojunctions, \cite{Chen_2006,Chen:2007,Chen2005QuantumSO}, in which Wigner crystallization occurs at large enough magnetic fields in between Quantum Hall plateaux.\footnote{The reader may question whether these systems are truly in a hydrodynamic regime. Here we observe that in fact a memory matrix analysis is enough to establish the expressions for the ac conductivities, \cite{Delacretaz:2017zxd}, so that the hydrodynamic assumption is not stricto censu needed, only that translations are weakly broken explicitly.} The conductivity is characterized by the peak frequency, $\omega_{pk}$, the magnetophonon damping rate $\Omega$ and an extra asymmetry parameter $a$ compared to the case without a magnetic field.
 Topological defects and pinning are expected to give independent contributions to the magnetophonon damping rate $\Omega$. These rates can be computed using the memory matrix formalism. For a defect-dominated phase, the ratio $\Omega/(\omega_{pk}a)=2$, while for a disorder-dominated phase where the magnetophonon predominantly relaxes into the electric current, $\Omega/(\omega_{pk}a)=1$. Interestingly, these values seem to account well for fits to the experimental results at low temperatures or strong magnetic fields. In the disorder-dominated case, the relaxation mechanism into a hydrodynamic current is reminiscent of the analogous mechanism in holographic systems discussed in sections \ref{section:HoloSp} and \ref{section:HoloSP}.

From the holographic perspective, the introduction of an external magnetic field in homogeneous models with broken translations has been considered in \cite{Baggioli:2020edn,Amoretti:2021lll,Donos:2021ueh}. A full holographic calculation of all linear transport coefficients together with matching to the hydrodynamic dispersion relation for the modes has not been performed yet.

%%%%%%%%%%%%%%%%%%%%%%%%%%%%%%%%%%%%%%%%%%%%%%

\section{Discussion: Transport in strange metals and pseudo-spontaneous breaking of translations}
\label{outlook:cuprates}

Can the physics of pseudo-spontaneous breaking of translations shed light on the phenomenology of high T$_c$ superconductors, in particular on their strange metallic phase? Transport experiments famously measure a resistivity linear in temperature \cite{PhysRevLett.59.1337} which extends for optimally doped samples from above room temperature to the lowest temperatures experimentally available when a magnetic field suppresses superconductivity. This observation brings two important puzzles. The absence of resistivity saturation at high temperatures violates the Mott-Ioffe-Regel bound \cite{Gunnarsson:2003,Hussey:2004} and precludes any notion of quasiparticle-based transport, calling for other descriptions of transport in systems with short-lived excitations \cite{Hartnoll:2014lpa}. Charge transport in conventional metals with long-lived quasiparticles is often analyzed with the Drude model. Applying this framework to the resistivity of strange metals identifies a `Planckian' scattering rate, \cite{Bruin804}, which on theoretical grounds can be argued to be the shortest relaxation timescale consistent with Heisenberg's uncertainty principle, \cite{sachdev2011quantum, Zaanen2004} -- see \cite{Hartnoll:2021} for a recent review on Planckian dissipation in metals and bounds on transport. 

At low temperatures, the persistence of a $T$-linear component to the resistivity over a range of dopings \cite{Cooper:2009,doi:10.1098/rsta.2010.0196,Hussey_2013,Legros2019,Putzke2021} clashes both with Fermi liquid predictions of a $T^2$ resistivity, which is only fully recovered beyond the superconducting dome for very overdoped samples, and with conventional expectations of transport in the vicinity of a quantum critical point, \cite{sachdev2011quantum}, where quantum critical behavior is not expected outside the quantum critical cone.

The slope of the $T$-linear resistivity appears to be of the same order of magnitude across different materials, \cite{Legros2019}, which hints at a universal mechanism underpinning this phenomenon. Extrapolations of the resistivity to zero temperature show that the disorder of the sample does not play an important role, with values of the residual resistivity varying sometimes over an order of magnitude or more across materials. Further evidence of disorder-independence comes from ion-irradiation experiments \cite{Rullier-Albenque:1995,Rullier-Albenque:1997,Rullier-Albenque:2000,Rullier-Albenque:2003}, which show that resistivity curves simply shift upwards when disorder is increased without any change in the slope of the $T$-linear component.  

Transport experiments also report a $T^2$ cotangent of the Hall angle, \cite{Chien:1991}, and a magnetoresistance linear in the magnetic field at large field over a range of dopings, \cite{Hayes:2014,Giraldo-Gallo:2018,Ayres:2021}. This is once again at odds with quasiparticle-based transport and the Boltzmann equation, which predicts that the resistivity and the Hall angle are controlled by the same transport timescale, and that the magnetoresistance is quadratic in field. Instead, the different temperature dependencies of the resistivity and Hall angle are often interpreted in a two-timescale scenario, \cite{Anderson_1991,Coleman:1996a,Coleman:1996b}. More generally, there is some experimental support for two sectors contributing to transport, one coherent and the other incoherent, \cite{Licciardello:2019,Ayres:2021,Culo:2021}. Transport experiments in overdoped cuprates are often analyzed using the Boltzmann equation. Angle-dependent magnetoresistance experiments allow to infer the quasiparticle scattering rates (with qualitatively diffent results for different materials, \cite{Abdel-Jawad2006,Grissonnanche2021}), but do not always allow to reproduce in-plane transport experiments, \cite{Ayres:2021}. Thus, even on the overdoped side, the validity of Boltzmann transport is not entirely obvious.

Turning to optics, the ac in-plane conductivity in the strange metal regime above the temperature at which superconductivity sets in is Drude-like, with a peak centered at zero frequency and a width of order $T$.  At higher temperatures, a number of compounds reveal a transfer of spectral weight and the zero frequency peak moves off-axis to a nonzero frequency, \cite{Hussey:2004,Delacretaz:2016ivq}. The ac conductivity also features an infrared contribution \cite{Uchida:1991,Quijada:1995,Quijada:1999} extending beyond the peak, which scales as $|\sigma(\omega)|\sim\omega^{-2/3}$, \cite{VanderMarel:2003,Hwang:2007}. This is weaker than the expected Drude scaling $|\sigma(\omega)|\sim\omega^{-1}$. This resonates with the two-component analysis of transport, however fits to optics data typically assume that the dc conductivity solely originates from the Drude component, ascribing a frequency dependence to the infrared component which vanishes as $\omega\to0$. Recently, \cite{vanHeumen:2022} vindicated this picture in a careful study of optics across a range of dopings in single-layer BSCCO. From a theoretical perspective, there is a tension between assuming a gapless, scaling contribution decaying as some power of frequency for frequencies $\omega\gtrsim T$, but which would not produce a corresponding decaying power of temperature in the regime $\omega\lesssim T$, as $\omega/T$ scaling would dictate and as seems to hold well experimentally, \cite{vanHeumen:2022,Michon:2022}. It would be interesting to investigate to what extent this constraint in fitting optics can be relaxed and cross-referenced to dc transport data.

These experimental facts pose an immediate conundrum when attempting to interpret them in the framework of a metal with slowly-relaxing momentum. The ac conductivity at not too high temperatures suggests a Drude analysis may work, but fails to account for the appearance of an off-axis peak at higher temperatures or for the infrared non-Drude contribution. The ac conductivity of a slowly-relaxing metal is given in \eqref{sigmaGamma}. If momentum relaxes weakly, $\Gamma$ is small compared to some parameter determining the scale at which other degrees of freedom start to be important, usually temperature. But this theoretical assumption contradicts the experimental observation that $\Gamma\sim\mathcal O(T)$. The dc conductivity should be dominated by the `coherent' contribution from the Drude peak, $\sigma_{coh}\sim n_o^2/(\chi_{\pi\pi}\Gamma)+\mathcal O(\Gamma^0)$. $\Gamma$ strongly depends on disorder strength, \cite{Hartnoll:2016apf}, and so this contradicts the experiments where disorder is varied by ion-irradiation referred to in the previous paragraph.

While experimentally difficult to establish, the notion of coherent and incoherent charge transport in a slowly-relaxing metal is easy to understand from a theoretical standpoint. All that is required is to give up Galilean invariance, which imposes that the electric current is equal to the momentum density -- thereby killing any incoherent contribution to transport. Doing so, new processes are allowed that conduct charge but do not drag momentum, and neatly encapsulated in appearance of the transport coefficient $\sigma_o$ in the dc conductivity \eqref{sigmaGamma}. These processes naturally appear in hydrodynamics \cite{Davison:2015taa}, memory matrix approaches \cite{Lucas:2015pxa,Hartnoll:2016apf} and in holographic models \cite{Davison:2015bea}.

Relaxing Galilean invariance is not enough though, as in a metal with slowly relaxing momentum such incoherent processes inevitably give contributions to transport (of order $\Gamma^0$) subleading compared to the coherent contribution (of order $1/\Gamma$). There are several avenues one can think of to suppress the coherent contribution  to transport: 
\begin{enumerate}
    \item[i)] suppress the Drude weight through some emergent particle-hole symmetry that would effectively set $n_o=0$; 
    \item[ii)] assume strong explicit breaking of translations;
    \item[iii)] more radically, require that $\chi_{\pi\pi}\to+\infty$, \cite{Else:2021};  
    \item[iv)] short-circuit the large contribution from slowly-relaxing momentum by assuming translations are spontaneously broken, \cite{Delacretaz:2016ivq}.
    \end{enumerate}

Strange metals arise in doped Mott insulators, which leads to disregard i) (in contrast to the example of graphene near the charge neutrality point). The ability to synthesize very clean samples with a low residual resistivity \cite{Giraldo-Gallo:2018} also works against ii). iii) was recently considered \cite{Else:2021}. There, the authors argue that strange metals arise in the vicinity of an ordered phase where the order parameter has the same symmetries as loop currents \cite{PhysRevLett.83.3538, PhysRevB.73.155113} and that this would lead to the divergence of all susceptibilities in the same symmetry sector. It is interesting to note that holographic checkerboards, \cite{Withers:2014sja,Donos:2015eew,Cai:2017qdz}, naturally feature such current loops intertwined with translation symmetry breaking thanks to the bulk Chern-Simons terms. Here we note that fluctuations of the loop current order parameter have been put forward as the origin of the $T$-linear resistivity in the strange metallic phase \cite{RevModPhys.92.031001} as fermions scattering off them have a marginal Fermi liquid-like self-energy, \cite{PhysRevLett.63.1996}. The Sachdev-Ye-Kitaev model \cite{Chowdhury:2021qpy}, where a large number of species $N$ of fermions is introduced together with random interactions, provides a consistent theoretical framework realizing the marginal Fermi liquid self-energy. The flavor randomness and large $N$ limit make the computation of transport properties tractable. In its simplest incarnation, the $T$-linear term in the resistivity is perturbatively small (see also previous works on non-Fermi liquids without flavor randomness, \cite{Hartnoll:2014gba,Patel:2014jfa}). Recently, random (in flavor and real space) Yukawa-type couplings  to a gapless boson\footnote{Theories of non-Fermi liquids where fermions couple to a gapless boson have a long history, see \cite{Lee:2017njh} for a review.} have been shown to give rise to a $T$-linear term which is $\mathcal O(1)$ in the strength of spatial disorder, \cite{Patel:2022gdh}.  The gapless boson represents the fluctuations of an order parameter, at zero or nonzero wavector.
This suggests that the interplay between disorder and order parameter fluctuations might play an important role in understanding strange metals. Whether this $T$-linear component can arise over a range of dopings remains an open question.

Let us then consider (iv) how pseudo-spontaneous translation symmetry breaking may shed light on transport in strange metals. Further motivation for this is found in recent X-ray scattering reports of charge density fluctuations across the phase diagram, \cite{Peng:2018,Arpaia:2019,Miao:2021,Lin:2021,Lee:2020,Ma:2021,Arpaia:2021,Tam:2021,Kawasaki:2021,Lee:2021generic}, rather than restricted to the underdoped regime as previous experiments suggested, \cite{keimer2015quantum}. Besides charge density fluctuations at high temperatures, long-ranged, or at least longer-ranged than their underdoped counterparts, CDWs have now been reported on three different materials, \cite{Peng:2018,Miao:2021,Tam:2021}. Thus the strange metallic phase at optimal doping appears to be the only region where static CDW order has not been discovered (yet). The ubiquitousness of high temperature charge modulations is backed up by numerical (Determinant Quantum Monte Carlo) studies of the Hubbard model which also report intertwined charge and spin stripes at optimal doping and in the overdoped regime, \cite{Huang:2022}. Theoretical arguments on the impact of fluctuating charge density wave order on strange metal transport have been given in \cite{PhysRevB.95.224511,Delacretaz:2016ivq,Delacretaz:2017zxd,Amoretti:2018tzw,Seibold:2021,Delacretaz:2021qqu} (see as well the earlier references \cite{Kivelson1998,Taillefer:2010,PhysRevLett.108.267001,PhysRevB.86.115138} where the emphasis is more on the underdoped range).

We first discuss the ac conductivity in a pinned crystal, \eqref{accondpinned}. It is straightforward to see that the frequency dependence deriving from this formula interpolates between a Drude-like peak centered at $\omega=0$ if pinning $q_o$ is sufficiently weak compared to the typical frequency scales set by $\Gamma$ and $\Omega$, and an off-axis peak once pinning becomes stronger. A precise inequality can be derived from \eqref{accondpinned}, asking when all maxima in $\textrm{Re}~\sigma(\omega)$ are for $\omega=0$ or complex frequencies. This is as far as effective approaches can take us, since to determine how the frequency dependence of the conductivity varies in any given system requires a microscopic calculation or an experimental measurement. In gauge-gravity duality models, the peak can remain off-axis at all temperatures in the ordered phase, \cite{Andrade:2017ghg,Andrade:2017cnc,Andrade:2018gqk,Andrade:2020hpu}, or interpolate between being on-axis and off-axis, \cite{Amoretti:2018tzw,Donos:2019tmo}. In spectroscopic experiments, whether an off-axis peak develops at high temperatures seems very material-dependent -- materials where this behavior is seen are compiled in \cite{Delacretaz:2016ivq}. In YBa$_2$Cu$_4$O$_8$, the ac conductivity interpolates from Drude-like to an off-axis peak upon Zn-disordering, \cite{PhysRevLett.81.2132}. This is in qualitative agreement with charge transport in the pseudo-spontaneous regime, since stronger disorder will lead to an increase in the pseudo-Goldstone mass $q_o$ and in the pinning frequency $\omega_o$. It would be interesting to better understand the effects of Zn-doping on pinning charge density wave fluctuations in scattering experiments, \cite{Suchanek:2010,Guguchia:2017,Lozano:2021}, especially in light of the results in \cite{Arpaia:2019}.

Turning now to dc transport, it is clear that by looking at the dc conductivity of a pinned crystal \eqref{dcconductivity} alone, it will be hard to disentangle the individual contributions of various scattering processes.\footnote{See \cite{Amoretti:2019buu} for an attempt at fitting the hydrodynamic theory of pinned charge density waves in a magnetic field to magnetotransport data in a cuprate. While this analysis has the merit of fitting a consistent set of data on a single material, the set of data used does not allow to unambiguously determine all the parameters in the effective theory.} This said, we can distinguish two types of processes: 
\begin{itemize}
    \item First, extrinsic processes, encapsulated in the momentum relaxation rate $\Gamma$. This is through this relaxation coefficient that e.g. disorder or umklapp processes feed in the dc conductivity. Their scaling is expected to be sensitive to irrelevant deformations and to the details of the disorder distribution, leading to scattering rates $\Gamma_{ext}\sim T(g/T^{\Delta_g})^2\ll T$, \cite{Hartnoll:2012rj,Lucas:2014zea,Davison:2018ofp}. For this reason, it is unlikely they are the origin of the $T$-linear resistivity.
    \item Second, intrinsic processes, coming from dissipation into the bath of thermal, critical excitations, encapsulated in transport coefficients such as $\sigma_o$, $\xi_\mu$ and $\xi$.

 These are much stronger candidates as the source of $T$-linear resistivity. Gauge-gravity duality allows to easily calculate these transport coefficients and verify that indeed they their temperature dependence reflects the scaling properties of the underlying critical phase, \cite{Davison:2015taa,Davison:2018ofp,Davison:2018nxm}. These results have inspired scaling theories to explain transport data in cuprates, such as \cite{Hartnoll:2015sea,Karch:2015zqd}. A crucial extra ingredient compared to previous attempts at a scaling theory (e.g. \cite{Phillips:2005}) is the introduction of anomalous scaling dimensions for the charge density at the critical point, \cite{Gouteraux:2013oca,Gouteraux:2014hca,Karch:2014mba,LaNave:2019mwv}.\footnote{ Holographic models combining explicit breaking of translations and these new scaling laws met with difficulties, \cite{Davison:2013txa,Blake:2014yla,Amoretti:2016cad,Blauvelt:2017koq}, including matching all scaling laws and or suppressing the coherent, extrinsic contribution to the conductivity from momentum without resorting to strong explicit breaking. This task is made harder by the experimental hurdle of producing thermoelectric transport data displaying clean scaling laws over sufficiently large ranges of temperature. How such scaling theories extend to pseudo-spontaneously broken translations has not been investigated.}
     \end{itemize}

As we have already emphasized in section \ref{section:chargetransport}, introducing pseudo-spontaneous breaking of translations short-circuits the extrinsic contribution to the resistivity, which is now $\rho_{dc}\sim \mathcal O(\Gamma^0)$ rather than $\mathcal O(1/\Gamma)$ in a metal. The order $O(\Gamma^0)$ terms are determined by $\sigma_o$, $\xi_\mu$ and $\xi$, are intrinsic and are dominant against the extrinsic $O(\Gamma)$ terms. From \eqref{accondpinned} and \eqref{dcconductivity}, it is clear that they contribute both to the coherent (the peak) and to the incoherent (the infrared band) parts of the conductivity.This gives further motivation to revisit the two-component analysis of ac conductivity data, which customarily assumes that the infrared band does not contribute to the dc conductivity.

Heavily overdoped, non-superconducting cuprates feature a purely $T^2$ resistivity, while the $T$-linear component turns on at the onset of superconductivity, turning gradually stronger until the critical doping where the resistivity is purely $T$-linear. CDW order has not been reported for non-superconducting samples, and so far does not extend all the way to the edge of the superconducting dome for all superconducting overdoped samples, \cite{Tam:2021}.  Plots of the derivative of the resistivity with respect to temperature in overdoped LSCO and Tl2201 show a gradual change of slope below about 250K from a high temperature, $T$-linear incoherent bad metallic behavior, to a low temperature $T+T^2$ behavior, \cite{doi:10.1098/rsta.2010.0196,Hussey_2013,Putzke2021}, for all overdoped samples including those where X-ray experiments do not find a static CDW, \cite{Tam:2021}. Whether the change in the resistivity slope can be more precisely connected to the onset of charge density fluctuations and static CDW order at lower temperatures remains to be clarified. 

{Bearing this caveat in mind, we can assume that the temperature dependence of $\Gamma\sim \gamma_0+\gamma_2 T^2+\ldots$, originating from disorder (the zero temperature residual resistivity) and umklapp (the Fermi liquid-like behavior recovered outside the superconducting dome). On the other hand, intrinsic processes controlling the CDW contribution relaxation to the resistivity might be responsible for the disorder-independent, $T$-linear component at low temperatures. The magnitude of this contribution is naturally proportional to the elastic modulus and would be expected to become stronger as temperature is decreased and the CDW order sets in, consistently with the increase in the onset temperature of the linear component as doping decreases. }

Why should those intrinsic processes carry a $T$-linear dependence? This is a difficult question, barring a concrete microscopic model of cuprates. The holography-inspired scaling theories alluded to above give one possible answer, but have not been extended to the pseudo-spontaneous case yet. 

An alternative relies on theoretical arguments by which diffusivities $D$ in strongly-correlated systems tend to saturate a Planckian bound, \cite{Hartnoll:2014lpa}, 
\begin{equation}
\label{Dbound}
    D\gtrsim v^2\tau_{Pl}\,,\quad \tau_{Pl}=\frac{\hbar}{k_B T}\,.
\end{equation}
Here $v$ is some characteristic velocity, which is sometimes argued to the Fermi velocity, the Lieb-Robinson velocity or the butterfly velocity (see e.g. \cite{Blake:2016wvh}). Through this general mechanism, applied to the diffusive transport coefficients $\sigma_o$, $\xi_\mu$ and $\xi$, we may expect various disorder-independent, $T$-linear contributions to the resistivity, split between the coherent and incoherent terms. This resonates with the analysis of the magnetoresistance data of \cite{Ayres:2021}, which found necessary to include a $T$-linear component in both coherent and incoherent contributions. 

The diffusivity $\sigma_o$ is directly related to the thermal diffusivity, \cite{Davison:2018ofp}. Energy diffusion is likely to be universal in a critical phase. Indeed, measurements of this observable in hole-, \cite{Kapitulnik}, and electron-doped cuprates, \cite{kap2}, as well as in crystalline insulators, \cite{Behnia_2019,Mousatov:2020}, all suggest that the thermal diffusivity in these materials is close to a Planckian bound. 

The reader may legitimately wonder why the same ought to hold for the Goldstone diffusive coefficient $\xi$. The Goldstones are weakly-coupled in the low energy effective field theory, \cite{Son:2002zn,Nicolis:2013lma,Nicolis:2015sra}, and so it does not naturally follow that they relax on Planckian scales (the attenuation of superfluid phonons being a case in point). On the other hand, in sections \ref{section:HoloSp}, \ref{section:HoloSP} and \ref{sec:4}, we have highlighted a dissipation mechanism into hydrodynamic currents at play both in holographic systems and in 2d electron gases hosting Wigner crystal phases. This mechanism links the Goldstone diffusivity $\xi$ to the thermal diffusivity, which itself is likely to be close to a Planckian bound in a strongly-correlated system.

At the critical doping, the resistivity is purely $T$-linear with an $O(1)$ coefficient. At this doping, quasiparticles are completely lost due to strong correlations, \cite{doi:10.1126/science.aar3394}, vindicating the applicability of quantum bounds on transport of the kind \eqref{Dbound}. In the absence of quasiparticles, the Goldstone sound velocity is a plausible candidate to enter in the bound, in which case the factors of the elastic moduli cancel out from the resistivity, yielding an $O(1)$ prefactor for the $T$-linear resistivity.

On the other hand, if the strange metal regime near critical doping is related to a kind of CDW critical point dominated by charge density fluctuations, then fluctuations of the amplitude of the order parameter ought to be included in the effective description, not just its phase, \cite{RevModPhys.49.435} (see \cite{Grossi:2021gqi} for a recent application to QCD in the chiral limit), bringing us back to the arguments developed in \cite{Patel:2022gdh} for the origin of the $T$-linear resistivity at critical doping. Holography will also certainly be a valuable tool to construct such EFTs augmented with order parameter fluctuations, \cite{Herzog:2010vz,Donos:2022xfd}.

Pseudo-spontaneous breaking of translations thus appears to be a promising avenue to understand various features of strange and bad metals. While it is difficult to be more conclusive at this stage, further analyses of experimental data, revolving around the influence of disorder on charge density fluctuations, a systematic analysis of charge, heat and magneto-transport data on the same compound, and a refinement of the two-component analysis of optics data, may give further support to this hypothesis or disprove it.

\begin{acknowledgments}
We would like to thank Aristomenis Donos for collaboration at an early stage of this work and for several fruitful discussions on the topics of this \textit{Colloquium}.
We would like to thank Luca Delacr\'etaz, Sean Hartnoll, Erik van Heumen, Akash Jain, Stephen Kivelson, John Tranquada, Jan Zaanen for numerous inspiring discussions on the topic of phases with spontaneously broken translations over the years. We would also like to thank all our collaborators for our common work, which provided part of the basis for this review. We also thank Sean Hartnoll, Akash Jain, Matti Järvinen, Alexander Krikun, Li Li and Vaios Ziogas for pointing out various misprints and providing helpful feedback on a previous version of this manuscript.
M.B. acknowledges the support of the  Shanghai Municipal Science and Technology Major Project (Grant No.2019SHZDZX01). The work of B.G. is supported by the European Research Council (ERC) under the European Union's Horizon 2020 research and innovation program (grant agreement No758759). 
\end{acknowledgments}

\appendix

\section{Positivity of entropy production \label{app:entropyproduction}}

In this Appendix, we give more details on the steps leading to the Lorentz invariant constitutive relations \eqref{constreldiss}. Using \eqref{constrelideal} together with the first law of thermodynamics
\begin{equation}
    Tds=d\varepsilon-\mu dn-v_i d\pi^i-h^{ij}d\left(\nabla_i u_j\right)
\end{equation}
as well as the equations of motion, the divergence of the entropy current is found to be
\begin{equation}
\label{entropydivergence}
\begin{split}
    T\dot s+T\nabla_i \left(\frac{j_q^i}T\right)=&\tilde u^j\left(K_j+\nabla^i h_{ij}\right)\\
    &-\tilde j_q^i\frac{\nabla_i T}T-\tilde j^i\nabla_i\mu-\tilde \tau^{ij}\nabla_i v_j\,,
    \end{split}
\end{equation}
with 
\begin{equation}
    j_q^i=Ts v^i+\tilde j_q^i\,,\quad  \tilde j_q^i=\tilde j_\varepsilon^i-\mu \tilde j^i+h_{ij}\tilde u^j-v_j\tilde\tau^{ij}\,.
\end{equation}
Here we have turned on an external source for the $u^i$, $f_{el}\mapsto f_{el}-K_i u^i$, which we take to be first order in gradients $K_i\sim\mathcal O(\nabla)$.

The right-hand side must be positive so that entropy is not destroyed by dissipative processes. This constrains the constitutive relations to take the following form:\footnote{The ideal equations of motion are used to remove all time derivatives in the constitutive relations, and we choose a frame such that the conserved densities are not corrected at first order in gradients. See \cite{Kovtun:2012rj,Kovtun:2019hdm} for a discussion on the role of frames in relativistic hydrodynamics, and \cite{deBoer:2017ing,deBoer:2017abi,Novak:2019wqg,Poovuttikul:2019ckt,deBoer:2020xlc,Armas:2020mpr} for hydrodynamics without boosts.}
\begin{equation}
\label{constreldissapp1}
\begin{split}
&\tilde j^i=-\sigma_o^{ij}\nabla_j\mu-\alpha_o^{ij}\nabla_j T-\gamma_\mu^{ij}\left(K_j+\nabla^k h_{kj}\right)\,,\\
&\frac{\tilde j_q^i}T=-\bar\alpha_o^{ij}\nabla_j \mu-\frac{\bar\kappa_o^{ij}}{T}\nabla_j T-\gamma_T^{ij}\left(K_j+\nabla^k h_{kj}\right)\,,\\
&\tilde\tau^{ij}=-\eta^{ijkl}\nabla_{(k}v_{l)}\,, \\
&\tilde u^i=\xi_{\mu}^{ij}\nabla_j\mu+\xi_T^{ij}\nabla_jT+\xi_h^{ij}\left(K_j+\nabla^k h_{kj}\right)\,.
\end{split}
\end{equation}
Turning on the external source $K_j$ is necessary to remove terms like $\nabla_j h^k_k$, which otherwise would appear to be allowed. In the main text and in the remainder of this Appendix, we now turn off the external sources.

The Onsager relations can be imposed either on the matrix of retarded Green's function
\begin{equation}\label{Onsager_GR}
 S \cdot\left(G^R(\omega,-q)\right)^T = G^R(\omega,q)\cdot S\,,
\end{equation} 
or, as is often simpler, directly on the $M\cdot \chi$ matrix
\begin{equation}\label{Onsager_Mchi}
S \cdot\left(M(-q)\cdot\chi\right)^T = M(q)\cdot\chi\cdot S\,,
\end{equation}
where $M$ is defined from the equations of motion and
with $S$ being the matrix of time-reversal eigenvalues of the corresponding fields $(n,\varepsilon,\pi_\parallel,\lambda_\parallel,\pi_\perp,\lambda_\perp)$. Here $S=\textrm{diag}(1,1,-1,1,-1,1)$.

The $M\cdot \chi$ matrix reads
\begin{equation}
	\label{MchiWC}
	M\cdot\chi=  \left(\begin{array}{cccccc}
		\sigma_0 q^2 &\alpha_0 q^2 &iq n &\gamma_{\mu}q^2 &0 &0 \\
		\bar\alpha_0 q^2 &\frac{\bar\kappa_0}{T} q^2 &iq s &\gamma_{T}q^2 &0 &0 \\
		iq n &iq s &(\zeta+\eta)q^2 &-iq &0 &0 \\
		\xi_\mu q^2 &\xi_T q^2 &-iq &\xi q^2 &0 &0 \\
		0 &0 &0 &0 &\eta q^2 &-iq  \\
		0 &0 &0 &0 &-iq &\xi q^2 
	\end{array}\right)\,.
\end{equation}
The Onsager relations further fix
\begin{equation}
\label{OnsagerRel}
    \gamma_{\mu}=\xi_\mu\,,\quad \gamma_{T}=\xi_T
\end{equation}
Recall that all the transport coefficient matrices and tensors are decomposed as e.g. $\sigma_{o}^{ij}=\sigma_{(o)}\delta^{ij}+\sigma_{(u)}u^{ij}$, and the final coefficient appearing in \eqref{MchiWC} is a linear combination of $\sigma_{(o)}$ and $\sigma_{(u)}$, for instance $\sigma_o=\sigma_{(o)}+(u_o/2)\sigma_{(u)}$.

At linearized level, it is enough for us to impose positivity of \eqref{MchiWC}, but in general, one should instead require the quadratic form on the right-hand side of \eqref{entropydivergence} to be positive definite. Positivity of \eqref{MchiWC} follows if all eigenvalues are positive, which in turn is equivalent to all principal minors of this matrix being positive. The following constraints are sufficient to that effect
\begin{equation}
\label{entropybounds}
\begin{split}
    &\sigma_o\,,\bar\kappa_o\,,\eta\,,\zeta+\eta\geq0\,,\\
    &\sigma_o\bar\kappa_o\geq T\alpha_o^2\,,\quad \sigma_o\xi\geq\xi_\mu^2\,,\quad\bar\kappa_o\xi\geq T\xi_T^2\,.
    \end{split}
\end{equation}

The Lorentz boost Ward identity implies that $j_\varepsilon^i=\pi^i$. At ideal level, using \eqref{constrelideal} this fixes 
\begin{equation}
\chi_{\pi\pi}=\varepsilon+p-p_{el}\,,
\end{equation}
while at first order in gradients, from \eqref{constreldissapp1} and \eqref{OnsagerRel}, the following relations between the longitudinal transport coefficients
\begin{equation}
    \label{Lorentzlimit}
    \begin{split}
    &T\xi_T+\mu\xi_\mu-p_{el}\xi=0\\
    &T\alpha_o+\mu\sigma_o-p_{el}\xi_\mu=0\\
    &\mu\alpha_o+\bar\kappa_o-p_{el}\xi_T=0\,,
    \end{split}
\end{equation}
or in matrix form:
\begin{equation}
      \begin{split}
      &\alpha_o^{ij}+\frac{\mu}{T}\sigma_o^{ij}+\frac1T h_{ik}\xi_\mu^{kj}=0\\
     & \kappa_o^{ij}+\frac{\mu}{T}\alpha_o^{ij}+\frac1T h_{ik}\xi_T^{kj}=0\\
      &\xi_T^{ij}+\frac{\mu}{T}\xi_\mu^{ij}+\frac1T h_{ik}\xi_h^{kj}=0\,.
      \end{split}
\end{equation}
The constitutive relations then become
\begin{equation}
\label{constreldissapp}
\begin{split}
&\tilde j^i=-T\sigma_o^{ij}\nabla_j\frac{\mu}{T}-\gamma_\mu^{ij}\nabla^k h_{kj}\,,\\
&\frac{\tilde j_q^i}T=\left(\mu\sigma_o^{ij}+ h^i{}^{l}\xi_\mu^{lj}\right)\nabla_j\frac{\mu}{T}-\left(\mu\xi_\mu^{ij}+ h^i{}_{l}\xi_h^{lj}\right)\nabla^k \frac{h_{kj}}T\,,\\
&\tilde\tau^{ij}=-\eta^{ijkl}\nabla_{(k}v_{l)}\,, \\
&\tilde u^i=T\xi_{\mu}^{ij}\nabla_j\frac{\mu}T+\xi_h^{ij}\nabla^k \frac{h_{kj}}T \,.
\end{split}
\end{equation}

In the Galilean limit, the Galilean boost Ward identity enforces $j^i\propto\pi^i$ and instead:
\begin{equation}
\label{Galileanlimit}
    \sigma_o^{ij}=0\,,\quad \alpha_o^{ij}=0\,,\quad \xi_\mu^{ij}=0\,.
\end{equation}
\vskip4cm

\bibliography{rmpbib}

\end{document}